\documentclass[aps,prl,twocolumn,showpacs,groupedaddress]{revtex4}
\usepackage{graphics}

\begin{document}

\title{Effects of Pauli paramagnetism on superconducting vortex 
phase diagram in strong fields}

\author{Hiroto Adachi and Ryusuke Ikeda}

\affiliation{%
Department of Physics, Kyoto University, Kyoto 606-8502, Japan
}

\date{\today}

\begin{abstract}

 The Ginzburg-Landau (GL) functional and the resultant phase diagram of 
quasi two-dimensional superconductors in strong fields perpendicular to the layers, 
where the Pauli paramagnetic depairing is not negligible, 
are examined in details by assuming the weak coupling BCS model with 
a $d_{x^2-y^2}$-like pairing. It is found that the temperature at which 
the mean field (MF) transition at $H_{c2}$ changes into a discontinuous one 
lies much above another temperature at which a line $H_{\rm FFLO}(T)$ of 
transition to a helical FFLO-like vortex solid may begin to appear. 
In addition to MF results, details of a real phase diagram near 
$H_{c2}(T)$-line are examined based on a theoretical argument and 
Monte Carlo simulations, and it is found that the MF discontinuous transition 
is changed due to the fluctuation into a crossover which is nearly 
discontinuous 
in systems with weak enough fluctuation. These results are 
consistent {\it both} 
with the MF discontinuous behavior and a suggestion of $H_{\rm {FFLO}}(T)$ 
in the heavy fermion superconductor CeCoIn$_5$ with weak fluctuation {\it and} 
with their absence in cuprate and organic superconductors with 
strong fluctuation. 

\end{abstract}


\maketitle

\section{I. Introduction}

Traditionally, effects of Pauli paramagnetism on superconductors with 
a spin-singlet Cooper-pairing 
have been discussed by simply focusing on two energy scales \cite{Clogston,
Tsuneto}; the superconducting (SC) condensation energy and 
the Zeeman energy preventing the singlet pairing. 
This is a reasonable explanation on the first order transition (FOT) 
in an exceptional case with no orbital depairing creating field-induced vortices, 
i.e., a thin-film superconductor in parallel fields \cite{Adams}. 
Further, there is also a possibility of a structural transition {\it within} 
the Meissner phase into the so-called FFLO state \cite{FF,LO} with a periodic 
modulation, induced by the spin (paramagnetic) depairing, of the SC order 
parameter. However, the orbital depairing effects, i.e., the field-induced vortices, 
are inevitably present in most of cases of a bulk type II superconductor under 
a strong field, including a layered material under a field parallel to 
the superconducting layers \cite{Isotani}. Hence, we encounter quite 
a {\it different} issue from that in the works \cite{Clogston,Tsuneto,FF,LO,Agterberg}, 
that is, effects of the spin depairing on the {\it vortex state} 
which has {\it no} Meissner response. Since the number of vortices is 
determined only by the magnetic field strength and system sizes, treating the orbital depairing 
as a perturbation for the case with no orbital depairing is not valid. 

At present, it is well understood \cite{Natter,Ikeda2} that, in lower fields where 
the spin depairing is negligible, the $H$-$T$ phase diagram for the vortex states 
is {\it drastically} changed by including the SC fluctuation neglected 
in the mean field (MF) approximation. A typical one among such drastic fluctuation 
effects is the fact that the second order MF transition at $H_{c2}$ is not 
realized as a consequence of the fluctuation and gives way to a {\it weak} 
first order transition lying below $H_{c2}$ between the vortex solid and the 
vortex liquid region which needs not be distinguished from the normal state. 
At least theoretically, it is important 
to extend this issue to the strong field 
region in which the spin depairing is not negligible and the MF transition 
at $H_{c2}$ may be discontinuous. 

Through previous MF works on the vortex states of superconductors with 
paramagnetic depairing \cite{GG,Houzet,Maki}, however, one notices that 
even the $H$-$T$ phase diagram in the MF approximation is an unsettled issue.
 For instance, a FOT induced by the spin depairing was expected through 
a calculation in dirty limit \cite{Maki}, to the best of our knowledge, 
contrary to the experimental facts. Further, even in pure (clean) limit, 
there are no consensus on the MF phase 
diagram. 
In strong fields of our interest, any Meissner phase 
(i.e., any phase occurring with no orbital depairing) is not possible, 
and we expect just some vortex solids, such as the ordinary solid consisting 
of straight vortex lines and an, if any, FFLO-like solid state with 
a periodic modulation along the applied field, as SC ground states 
in clean limit with no defects leading to a vortex-pinning. 
Let us call a transition curve between the above-mentioned two vortex solid states 
as $H_{\rm FFLO}(T)$. In ref.\cite{GG}, the transition at $H_{\rm FFLO}$ 
was argued to be of first order with no detailed calculation, while it was 
obtained as a second order one in ref.\cite{Houzet} where the orbital 
depairing 
represented by the gauge-invariant gradient is treated perturbatively. 
Further, the temperature $T^*$ at which the MF transition at $H_{c2}$ changes 
into a discontinuous one was concluded there \cite{Houzet} to lie much below 
another temperature $T_{\rm FFLO}$ at which $H_{\rm FFLO}(T)$ and $H_{c2}(T)$ branch. 
In addition, the $H_{\rm FFLO}(T)$-line is often suggested to be insensitive to $T$. 
We note that all of conflicting results raised above were obtained in the same model, 
i.e., the simplest weak-coupling BCS model. 

In the present paper, we consider the SC phase diagram of quasi two-dimensional 
(2D) superconductors with a singlet $d$-wave pairing under a high field 
perpendicular to the SC layers where both the orbital- and spin-depairings are not 
negligible. In Sec.II, a MF calculation for the weak-coupling BCS model is first 
performed by focusing primarily on the region where the SC order parameter is 
described within the lowest Landau level (LLL or $N=0$ LL). 
We demonstrate by treating the two depairing effects on the same footing that 
most of the previous MF results mentioned above may be significantly changed. 
Our analysis is different from that in ref.\cite{Houzet} in that the gauge-invariant 
gradient is fully incorporated. This nonperturbative inclusion of the orbital 
depairing should become important upon cooling and with increasing field, although, 
instead, multiple numerical integrals have to be performed to accomplish it. 
Further, bearing a comparison with data in real systems in mind, 
a weak impurity scattering should be incorporated because the $H_{\rm FFLO}(T)$ line 
which may appear in high $H$ and low $T$ is found to be quite sensitive to the 
impurity strength. 
Consequently, the MF phase diagram is determined by a competition among 
the two depairing effects and the impurity strength. 
We find that the temperature $T^*$ always lies above other two characteristic 
temperatures $T_{\rm FFLO}$ and $T_{\rm next}$ below which the vortex solid 
will be described by the next lowest Landau level ($N=1$ LL). In addition to 
these studies of the MF phase diagram, linear response and elastic properties 
in a possible helical FFLO-like vortex solid will be examined to give a result 
useful for search of such a helical solid state. 

Next, in Sec.III, we address effects of the SC fluctuation 
on the phase diagram 
by neglecting vortex pinning effects. 
We notice that a theoretical consideration on the fluctuation effects on the 
orderings in vortex states below $H_{c2}$ is unaffected by a change of the GL model 
brought by the presence of the spin depairing and naturally leads to the conjecture 
that the FOT obtained in the MF approximation (MF-FOT) at $H_{c2}$ will not 
occur as a true FOT in real systems. Actually, otherwise, the high field portion 
of the $H$-$T$ phase diagram would not become compatible with its low field 
portion in which the absence of the second order MF transition at $H_{c2}$ is 
well-established~\cite{Natter, Ikeda2}. 
The only transition in the case with no pinning disorder should be 
the melting transition of a vortex solid. In the present case with a MF-FOT, 
however, the GL model needs to have higher order (nonlinear) terms other than the quartic term, 
and this fact makes an analytic study more involved. For this reason, we have 
chosen to perform Monte Carlo simulations on a GL model justified through our 
microscopic analysis in order to examine the true phase diagram. 
Our simulation results are consistent with the above-mentioned conjecture that 
the MF-FOT at $H_{c2}$ is changed due to the fluctuation into a crossover of which 
the width is narrower in systems with weaker fluctuation. Together with the 
impurity-induced disappearance of the MF-FOT, 
this fluctuation-induced broadening of 
the sharp behavior reflecting the MF-FOT explains why 
the nearly discontinuous behavior at $H_{c2}$ has not been observed so far in, 
except the recent observations in CeCoIn$_5$ 
\cite{Movshovich,Izawa,Tayama,Bianchi,Murphy}, most of bulk type II superconductors 
with a spin-singlet pairing. On the other hand, we find that, 
in spite of the absence of the genuine FOT at $H_{c2}$, a hysteresis may appear 
near $H_{c2}$ in simulations for cases with weaker fluctuation 
(or equivalently, at very low temperatures) as a result of an incomplete 
relaxation at finite time scales. 

Finally in Sec.IV, experimental facts in CeCoIn$_5$ under ${\bf H} \parallel c$ 
and other materials are discussed based on the results obtained in Sec.II
and III. It is conjectured there that the SC fluctuation makes the 
$H_{\rm FFLO}(T)$-line and the FOT-like behavior at $H_{c2}$ unobservable in 
strongly fluctuating systems such as organic materials. Further, it is also 
pointed out there that, although microscopic results in 
Sec.II are not 
applicable to the ${\bf H} \perp c$ case in CeCoIn$_5$, recent specific heat 
data \cite{Movshovich,Radovan} can be understood within our results in Sec.III. 

We use the unit $\hbar=c=k_{\rm B}=1$ throughout the manuscript. 

\section{II. Mean Field Analysis} 
Following Klemm {\it et al.}\cite{Klemm}, 
we start from a BCS hamiltonian for a quasi 2D superconductor under 
a nonzero magnetic field perpendicular to the basal plane
\begin{eqnarray} 
{\cal H}&=& {\cal H}_0 + {\cal H}_J- \frac{|g|}{2} \sum_{\sigma, j} 
 \int \frac{d^2 q_{\perp}}{(2 \pi)^2} 
	B_{j}^{\sigma \dagger}({\bf q_\perp})
          B_{j}^{\sigma}({\bf q_{\perp}}).  
\label{BCS}
\end{eqnarray}
 Here, $g$ is the attractive interaction strength,  
$\varphi_j^{\sigma}({\bf r_\perp})=(1/\sqrt{\Omega s}) \sum_{\bf p_\perp} 
a_j^\sigma({\bf p_\perp}) e^{{\rm i}{\bf p_\perp \cdot r_\perp}}$ 
is the annihilation operator of 
a quasi-particle with spin $\sigma$ ($=+1$ or $-1$) 
at the in-plane position ${\bf r_\perp}$ on the $j$-th plane, 
and $s$ and $\Omega$ are the interlayer spacing and the area of a layer, respectively.  
 The pair-field operator is defined by 
$B_{j}^{\sigma}({\bf q_{\perp}}) = \sum_{\bf p_\perp} w_{\bf p}
a_j^{-\sigma}({\bf -p_\perp^-}) a_j^{\sigma}({\bf p_\perp^+})$, 
where $w_{\bf p}$ is the orbital part of the pairing-function and, 
in the case of $d_{x^2-y^2}$-pairing, is written as 
$\sqrt{2}({\hat p}_x^2 - {\hat p}_y^2)$, 
where ${\hat {\bf p}}$ is the unit vector in the ${\bf p}_\perp$ direction, 
and ${\bf p_{\perp}^\pm}$ implies ${\bf p_\perp \pm q}/2$.
 The first term ${\cal H}_0$ in eq.(\ref{BCS}) represents in-plane motions of 
quasi-particles, 
\begin{eqnarray}
{\cal H}_0 &=& s \sum_{\sigma, j}  \int d^2 r_{\perp} \Bigg(
        {\varphi}_{j}^{\sigma \dagger}({\bf r_\perp})
        \frac{ ({\rm -i}{\nabla}_{\perp} + |e|{\bf A} )^2}{2m} 
	{\varphi}_{j}^{\sigma}({\bf r_\perp}) \nonumber \\
&& \hspace{1cm} +  {\varphi}_{j}^{\sigma \dagger}({\bf r_\perp}) 
        \Big[ v_j({\bf r_\perp})-\sigma I \Big] {\varphi}_{j}^{\sigma}({\bf r_\perp}) 
     \Bigg), 
\label{intralayer}
\end{eqnarray} 
where the gauge field ${\bf A}$ has no out-of-plane component, 
$I=\mu_0 H$ is Zeeman energy, and 
$m$ is the effective mass of a quasi-particle. 
The strength of the paramagnetic depairing is measured by 
$\mu_0 H_{c2}^{\rm {orb}}(0)/2 \pi T_{c0}$ 
corresponding to the Maki parameter 
except a numerical factor, where $H_{c2}^{\rm {orb}}(T)$ is the MF transition curve 
in the case neglecting both the paramagnetic depairing and impurity effects, 
and $T_{c0}$ is the zero field ($H=0$) transition temperature in the pure 
limit. Throughout the numerical calculations in this and next sections, we will choose $\mu_0 H_{c2}^{\rm {orb}}(0)/2 \pi T_{c0}$ to be 0.8 by bearing a comparison with observations in CeCoIn$_5$ in mind (see sec.IV). The random potential 
$v_j({\bf r_\perp})$ obeys the following Gaussian ensemble: 
$\overline{ v_j({\bf r_\perp})}=0$, 
$\overline{ v_j({\bf r_\perp})v_{j'}({\bf r'_\perp})} 
= \delta({\bf r_{\perp}-r'_{\perp}}) \delta_{jj'}/2\pi N(0) \tau$, 
where $N(0)$ is a 2D density of state at the Fermi surface, and $\tau^{-1}$ 
the elastic scattering rate. 
 The second term represents the inter-plane hopping,  
\begin{eqnarray}
{\cal H}_J &=& \frac{Js}{2} \sum_{\sigma, j}  \int d^2 r_{\perp} \Bigg(
{\varphi}_{j}^{\sigma \dagger}({\bf r_\perp})
{\varphi}_{j+1}^{\sigma}({\bf r_\perp}) + {\rm H. c.} \Bigg),
\label{interlayer}
\end{eqnarray}
where H.c. denotes Hermitian conjugate. 
We use the familiar quasi-classical approximation for the single-particle propagator
\begin{equation}
G_{\varepsilon_\sigma}^H({\bf r,r'}) 
= G_{\varepsilon_\sigma}^{H=0}({\bf r-r'}) 
  e^{{\rm i} |e|\int_{\bf r}^{\bf r'} d{\bf s \cdot A} }.
\label{Green1}
\end{equation}
Here, the Green's function in zero field, $G_{\varepsilon_\sigma}^{H=0}({\bf r})$, 
is given as the Fourier transform of the expression 
\begin{equation}
G_{\varepsilon_\sigma}({\bf p})
=\Big[{\rm i}\tilde{\varepsilon}_\sigma-
\big(\xi_{{\bf p}_\perp}+J\cos(p_z s)\big) \Big]^{-1} ,
\label{Green2}
\end{equation}
where 
$\varepsilon$ denotes the Matsubara frequency $\varepsilon_n = 2 \pi T (n+1/2)$ 
in this paper, 
$\tilde{\varepsilon}_\sigma=\varepsilon+ s_\varepsilon/(2\tau)
-{\rm i}\sigma I $, and $s_\varepsilon= {\rm sgn}(\varepsilon)$. 
Since we take account of the paramagnetic depairing suppressing the 
MF upper critical field, 
the use of the quasi-classical approximation, valid if $k_F r_H \gg 1$, 
needs not to be questioned, where $k_F$ is the Fermi momentum, 
and $r_H=(2|e| H)^{-1/2}$ is the magnetic length. 
 According to the familiar Stratonovich-Hubbard procedure 
to introduce the pair-field $\Delta$, 
we can construct a GL functional. Throughout this paper, 
we neglect the {\it internal} gauge fluctuation and examine 
the resulting GL model in type II limit. 
The validity of this approximation will be explained in discussing 
the MF phase diagram later.

\subsection{A. Quadratic term}
\begin{figure}[t]
\scalebox{0.5}[0.5]{\includegraphics{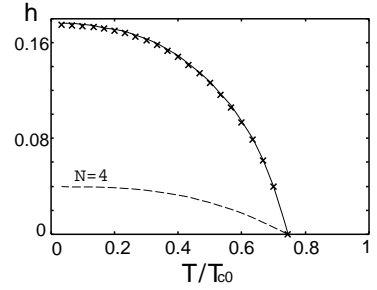}}
\caption{Comparison between $H_{c2}(T)$-curves for $(2 \pi T_{c0} \tau)^{-1}=0.1$ 
determined from $a_0(0)=0$ by neglecting (crossed symbols) and including (solid curve) 
the impurity-ladder vertex correction. 
The dashed curve is the line on which $a_4(0)=0$, and $h=H/H_{c2}^{\rm orb}(0)$.}
\label{fig:Hc2_VC}
\end{figure}

The quadratic term of the GL functional is given by 
\begin{equation}
{\cal F}_2 
= 	\sum_{q_z} \int d^2 r_{\perp} \tilde{\Delta}^*_{q_z}({\bf r_\perp})
	\left( \frac{1}{|g|}-\hat{K}_{2}({\bf \Pi}, q_z) \right) 
	\tilde{\Delta}_{q_z}({\bf r_\perp}),
\label{F21}
\end{equation}
where 
$\Delta_j = (N_{\rm layer})^{-1/2} \sum_{q_z} \tilde{\Delta}_{q_z}e^{{\rm i}q_z s j}$ 
denotes the pair-field on the $j$-th SC plane, 
$N_{\rm layer}$ is the number of SC planes, 
${\bf \Pi}=-{\rm i} \nabla_{\perp}+2 |e|{\bf A}$, 
and the operator $\hat{K}_2$ in the pure limit ($\tau^{-1} \to 0$) is simply given by 
\begin{eqnarray}
 \hat{K}_2 ({\bf \Pi}, q_z)
&=&
 2\pi N(0) T\sum_{\varepsilon > 0} \hat{D}_d(2\varepsilon), 
\nonumber \\
 \hat{D}_d(2\varepsilon) &=& 
\int_{\bf p} 
\frac{ \left< |w_{\bf p}|^2 G_{\varepsilon_\sigma}({\bf p})
 G_{-\varepsilon_\sigma}({\bf \Pi}+q_z \hat{\bf z}-{\bf p})\right>_\sigma }
{\pi N(0)}. 
\nonumber \\ \label{K21}
\end{eqnarray}
Here, the notation 
$\int_{\bf p} \equiv \int \frac{d^2 p_{\perp}}{(2 \pi)^2} \oint \frac{d(p_z s)}{2 \pi}$
was used. 
Hereafter, a bracket $\langle \cdots \rangle_{\alpha}$ with subscript $\alpha$ means 
an average over $\alpha$, so that $\langle \cdots \rangle_{\sigma}$ implies 
$2^{-1} \sum_{\sigma}$. 
In the impure case, there are also contributions from the impurity-ladder 
vertex corrections, and ${\hat D}_d(2\varepsilon)$ should be replaced as follows: 
\begin{eqnarray}
\hat{D}_d(2\varepsilon) &\to& 
\hat{D}_d(2\varepsilon+1/\tau) \nonumber \\
&&+ 
\left.\hat{E}_d(2\varepsilon)\frac{\Gamma}{1-\Gamma \hat{D}_s(2\varepsilon)}
\hat{E}_d(2\varepsilon)\right|_{2\varepsilon \to 2\varepsilon+1/\tau}, 
\label{Dd}
\end{eqnarray}
where $\Gamma=(2 \tau)^{-1}$, and $\hat{D}_s$ and $\hat{E}_d$ are defined by 
replacing $|w_{\bf p}|^2$ in eq.(\ref{K21}) into $1$ and $w_{\bf p}$ 
respectively. 
In the dirty limit with $2 \pi \tau T_{c0} \ll 1$, the impurity-ladder 
vertex correction expressed by the second term of eq.(\ref{Dd}) becomes important, 
although the SC phase itself in the present $d$-wave case is simultaneously suppressed. 
As is shown later, however, the MF-FOT appears only when $2 \pi T_{c0} \tau > 10$, 
and hence, we focus here on this moderately clean case. In fact, as given in 
Fig.\ref{fig:Hc2_VC}, 
a change in the $H_{c2}(T)$-curve due to the inclusion of the impurity-ladder vertex 
corrections is extremely small even when $(2 \pi T_{c0} \tau)^{-1} = 0.1$. 
Based on this result, the second term of eq.(\ref{Dd}) and any contribution 
including the impurity-ladder vertex corrections in other terms of 
GL functional will be neglected hereafter in the text and in Appendices. 

We should note here that, strictly speaking, the eigenstates of the operator 
$\hat{K}_2$ in the $d_{x^2-y^2}$-wave pairing are not the LLs, 
and that there are nonvanishing off-diagonal matrix elements between LLL and 
higher LLs with indices of multiples of four. 
However, in the range of Maki parameter considered in this paper, 
the instability line for the $N=4$ LL modes, defined by $a_4(0)=0$ in our notation 
used below, lies far below $H_{c2}(T)$ (see the dashed line in Fig.\ref{fig:Hc2_VC})
, and hence, 
we can neglect the off-diagonal elements in considering 
vortex states in $N=0$ (and $1$). 
Then, $w_{\bf p}$ in eq.(\ref{K21}) may be replaced by $1$, 
and our analysis using the LL basis becomes straightforward. 
When focusing on a projection $\tilde{\Delta}_{q_z}^{(N)}$ onto the $N$-th LL 
of $\tilde{\Delta}_{q_z} ({\bf r}_\perp)$, 
the corresponding eigenvalue of $\hat{K}_2$ is given by 
\begin{eqnarray}
 K_{2,N}(q_z) &=& 2 \pi T N(0) 
\int_{0}^{\infty} d \rho
f(\rho)
{\cal L}_N \Big( \frac{\rho^2}{2\tau_H^2} \Big) \nonumber \\
&&\quad \times   \exp{( -\rho^2/4\tau_H^2 )} 
{\cal J}_0 \Big(2J\sin(\frac{q_z s}{2})\rho\Big), \nonumber \\
\label{K22}  
\end{eqnarray}
where ${\cal L}_N$ is the N-th Laguerre function, ${\cal J}_0$ is the 
zeroth Bessel function, $\tau_H=m r_H/k_F$, 
and the function $f$ is defined by 
\begin{equation} 
f(\rho)=\frac{e^{-\rho/\tau} \cos( 2I\rho) }{\sinh(2 \pi T \rho)}. \label{f}
\end{equation}
The procedures leading to eq.(\ref{K22}) will be explained in Appendix A. 
After, like in $H=0$ case, eliminating the high energy cut-off by defining $T_{c0}$, 
we obtain the final expression for the quadratic free energy 
\begin{equation}
{\cal F}_2 = N(0)\sum_{N=0}^{\infty} \sum_{q_z} \int d^2 r_{\perp}
	a_N(q_z^2) \, 
	|\tilde{\Delta}_{q_z}^{(N)}({\bf r}_{\perp})|^2,
\label{F22}
\end{equation}
where
\begin{eqnarray}
a_N(q_z^2) &=& \ln(T/T_{c0}) + 
2 \pi T \int_{0}^{\infty} d \rho 
\Bigg( \big(\sinh(2 \pi T \rho) \big)^{-1} \nonumber \\
&-& f(\rho) {\cal L}_N \Big( \frac{\rho^2}{2\tau_H^2}\Big)
e^{- \big(\frac{\rho}{2 \tau_H} \big)^2} 
{\cal J}_0 \Big(2J\sin(\frac{q_z s}{2})\rho\Big) \Bigg). \nonumber \\
\label{aN1}
\end{eqnarray} 
For several purposes, it is convenient to expand $a_N(q_z^2)$ in powers of 
$\sin^2(q_z s/2)$:
\begin{equation}
a_N(q_z^2)=a_N(0) + a_N^{(1)} Q^2 + a_N^{(2)} Q^4 + \cdot \cdot \cdot , 
\label{aN2}
\end{equation}
where $Q \equiv J {\rm {sin}}(q_z s/2)$. 
When the SC transition in the MF approximation is of first order 
and occurs within the $N$-th LL, this MF-FOT line, which is a part of $H_{c2}(T)$, 
lies in the region where $a_N(0) > 0$, 
and the quartic and sixth order GL terms have to be considered to 
determine the $H_{c2}(T)$-line. 
A possibility of an instability to an FFLO-like helical vortex solid in the $N$-th LL 
may be studied, at least when $|a_N(0)| \ll 1$ (see discussions below 
eq.(\ref{flu})), by focusing on $a_N^{(1)}$ and $a_N^{(2)}$, where  
\begin{eqnarray}
a_N^{(1)} &=&  2\pi T \int_{0}^{\infty} d \rho 
 \rho^2
 f(\rho)    e^{- \big( \frac{\rho}{2\tau_H} \big)^2 } 
 {\cal L}_N \Big( \frac{\rho^2}{2\tau_H^2} \Big), \hspace{1cm}\\
a_N^{(2)} &=& - \frac{\pi T}{2} \int_{0}^{\infty} d \rho 
 \rho^4
 f(\rho)    e^{- \big( \frac{\rho}{2\tau_H} \big)^2 } 
 {\cal L}_N \Big( \frac{\rho^2}{2\tau_H^2} \Big). 
\end{eqnarray}
As far as $a_N^{(2)} > 0$, the instability point to an FFLO solid state is 
determined by $a_N^{(1)}=0$, and the corresponding transition is of second 
order. 

\begin{figure}[b]
\scalebox{1.0}[1.0]{\includegraphics{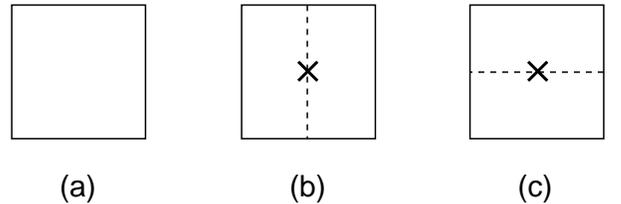}}
\caption{ Diagrams expressing the quartic term of the GL functional. 
The solid line implies the Green's function $G$, and the impurity (dashed) line 
carries $(2 \pi N(0) \tau)^{-1}$.} \label{fig:quart}
\end{figure}

\subsection{B. Quartic term}
The corresponding analysis for higher order (quartic and 6th-order) terms of 
the GL functional is more complicated than that for the quadratic one. 
As already explained, the impurity-ladder vertex corrections will be neglected 
in the ensuing analysis. 
Hereafter, it is convenient to work in a fixed Landau gauge ${\bf A}=(0, Hx, 0)$ 
and to represent the pair-field in the $N$-th LL in terms of the corresponding LL 
orbitals $u_{N,k}({\bf r}_\perp)$
\begin{equation}
\Delta_j^{(N)}({\bf r_\perp}) =  \frac{1}{\sqrt{S_H}}\sum_{k} 
 \phi_{N,k,j} \,\,  u_{N,k}({\bf r_\perp}), 
\label{LLexp}
\end{equation}
where ${S_H}={r_H L_y \pi^{1/2}}$. 
In the present gauge, $u_{N,k}({\bf r}_\perp)$ is given by 
\begin{equation}
u_{N,k}({\bf r_\perp}) = \frac{\hat{\pi}_+^N}{\sqrt{N !}} 
e^{-\frac{1}{2r_H^2}(x + kr_H^2)^2 +iky}, 
\label{eigenf} 
\end{equation}
where we introduce the creation and annihilation operators of LLs
\begin{equation}
\hat{\pi}_{\pm} = \frac{r_H}{\sqrt{2}}(\Pi_x \pm {\rm i} \Pi_y).
\label{LLope}
\end{equation}
Hereafter, let us focus on a vortex solid within the LLL-subspace. 
The corresponding analysis in higher LLs will be given elsewhere. 
Then, the quartic term of the GL functional can be written as 

\vspace{2mm}

\begin{widetext}
\begin{eqnarray}
{\cal F}_4 &=& 
\left. \frac{1}{2}\sum_j \int d^2 r_{\perp}
\hat{K}_4( \{ {\bf \Pi}_i \})
(\Delta_j^{(0)}({\bf r}_{\perp 1}) \Delta_j^{(0)}({\bf r}_{\perp 3}))^* 
\Delta_j^{(0)}({\bf r}_{\perp 2})
\Delta_j^{(0)}({\bf r}_{\perp 4})
\right|_{{\bf r}_{\perp i} \to {\bf r}_\perp}\\
\label{F41}
&=&
\left. \frac{1}{2S_H} \sum_{j,\{k_i\}} 
\phi^*_{0,k_1,j} \phi_{0,k_2,j} \phi^*_{0,k_3,j} \phi_{0,k_4,j}
\int \frac{ d^2 r_{\perp}}{S_H} \hat{K}_4( \{ {\bf \Pi}_i \}) 
u_{0k_1}^*({\bf r}_{\perp 1})u_{0k_2}({\bf r}_{\perp 2})
u_{0k_3}^*({\bf r}_{\perp 3})u_{0k_4}({\bf r}_{\perp 4})
\right|_{{\bf r}_{\perp i} \to {\bf r}_\perp},
\label{F42}
\end{eqnarray}
where ${\bf \Pi}_i = {\bf \Pi}({\bf r}_i)$. 
 In the impure case $\hat{K}_4$ consists of three terms 
represented in Fig.\ref{fig:quart} and will be expressed as 
$\hat{K}_4=\hat{K}_{\rm 4a}+\hat{K}_{\rm 4bc}$. 
The term $\hat{K}_{\rm 4a}$ is given by 
\begin{eqnarray}
\hat{K}_{4\rm a} &=&
2\pi T N(0)\sum_{\varepsilon}\left< 
\frac{-{\rm i} s_{\varepsilon} |w_{\bf p}|^4 }
{(2{\rm i}\tilde{\varepsilon}_\sigma + {\bf v \cdot \Pi}_1^*)
(2{\rm i}\tilde{\varepsilon}_\sigma + {\bf v \cdot \Pi}_2)
(2{\rm i}\tilde{\varepsilon}_\sigma + {\bf v \cdot \Pi}_3^*)
}
\right>_{\sigma,{\bf \hat{p}}}
+({\bf \Pi}_2 \leftrightarrow {\bf \Pi}_4)\\
\label{K4a1}
&=&
2\pi T N(0) \int \prod_{i=1}^3 d\rho_i 
f(\sum_{i=1}^3 \rho )
\left< |w_{\bf p}|^4 e^{{\rm i}(\rho_1 {\bf v \cdot \Pi}_1^* 
+\rho_2 {\bf v \cdot \Pi}_2+\rho_3 {\bf v \cdot \Pi}_3^* )} \right>_{\bf \hat{p}}
+({\bf \Pi}_2 \leftrightarrow {\bf \Pi}_4), 
\label{K4a2}
\end{eqnarray}
where ${\bf v}=k_F {\hat p}$/m, the function $f$ is defined by eq.(\ref{f}), 
and the bracket $\left< \,\,\, \right>_{\bf \hat{p}}$ implies the angle-average 
over the unit vector ${\bf \hat{p}}={\bf p}/k_F$ on Fermi surface. 
 The sum of Fig. \ref{fig:quart} (b) and (c), $\hat{K}_{\rm 4bc}$, is given by 
\begin{eqnarray}
\hat{K}_{4\rm bc} &=& -
\frac{2\pi T}{\tau} N(0)\sum_{\varepsilon}
\left<
\left< \frac{|w_{\bf p}|^2}
{(2{\rm i}\tilde{\varepsilon}_\sigma + {\bf v \cdot \Pi}_1^*)
(2{\rm i}\tilde{\varepsilon}_\sigma + {\bf v \cdot \Pi}_2)} \right>_{\bf \hat{p}}
\left< \frac{|w_{\bf p'}|^2}{(2{\rm i}\tilde{\varepsilon}_\sigma 
+ {\bf v' \cdot \Pi}_3^*)
(2{\rm i}\tilde{\varepsilon}_\sigma + {\bf v' \cdot \Pi}_4)}
 \right>_{{\bf \hat{p}'}}\right>_{\sigma}\nonumber \\
&&\hspace{12cm}+({\bf \Pi}_2 \leftrightarrow {\bf \Pi}_4) \\
\label{K4bc1}
&=&
-\frac{2\pi T}{\tau} N(0) \int \prod_{i=1}^4 d\rho_i 
f(\sum_{i=1}^4 \rho )
\left< |w_{\bf p}|^2 e^{{\rm i}(\rho_1  {\bf v \cdot \Pi}_1^* 
+ \rho_2 {\bf v \cdot \Pi}_2 )} 
\right>_{\bf \hat{p}}
\left< |w_{\bf p'}|^2 e^{{\rm i}(\rho_3 {\bf v' \cdot \Pi}_3^* 
+\rho_4 {\bf v' \cdot \Pi}_4 )} 
\right>_{\bf \hat{p}'}
+({\bf \Pi}_2 \leftrightarrow {\bf \Pi}_4),
\label{K4bc2}
\end{eqnarray}
where ${\bf v}'=k_F {\hat {\bf p}}'/m$. 
The following formulas which are derived in Appendix B are quite convenient. 
\begin{eqnarray}
e^{{\rm i} \rho {\bf v \cdot \Pi}} u_{0,k}({\bf r}_\perp)
&=&
e^{-\frac{(|\lambda|^2-\lambda^2)}{4}}
e^{-\frac{1}{2}(x/r_H+kr_H+\lambda )^2+{\rm i}ky}, \\
e^{{\rm i} \rho {\bf v \cdot \Pi}^*} u_{0,k}({\bf r}_\perp)
&=&
e^{-\frac{(|\lambda|^2-\lambda^{*2})}{4}}
e^{-\frac{1}{2}(x/r_H+kr_H - \lambda^* )^2-{\rm i}ky}.
\label{eigenope}
\end{eqnarray}
where $\lambda=\rho \zeta^*/\tau_H$ and 
$\zeta=\hat{p}_x + {\rm i} \hat{p}_y$ 
is the complex coordinate. Using this identity and eq.(\ref{eq:I4}) in Appendix B, 
we obtain the following results
\begin{eqnarray} 
&&\left. 
\int \frac{d^2 r_{\perp}}{S_H} \hat{K}_{4a}( \{ {\bf \Pi}_i \}) 
u_{0k_1}^*({\bf r}_{\perp 1})u_{0k_2}({\bf r}_{\perp 2})
u_{0k_3}^*({\bf r}_{\perp 3})u_{0k_4}({\bf r}_{\perp 4})
\right|_{{\bf r}_{\perp i} \to {\bf r}_\perp} \nonumber \\
&&=
\frac{2\pi T N(0)}{\sqrt{2}} \int \prod_{i=1}^3 d\rho_i 
f(\sum_{i=1}^3 \rho )
\int \frac{d^2 r_\perp}{S_H}
\Bigg< 
|w_{\bf p}|^4
e^{{\rm i}(\rho_1 {\bf v \cdot \Pi}_1^* 
+\rho_2 {\bf v \cdot \Pi}_2+\rho_3 {\bf v \cdot \Pi}_3^* )} \nonumber \\
&& \hspace{7cm}\times
u_{0k_1}^*({\bf r}_{\perp 1})u_{0k_2}({\bf r}_{\perp 2})
u_{0k_3}^*({\bf r}_{\perp 3})u_{0k_4}({\bf r}_{\perp 4})
\Bigg>_{\bf \hat{p}}
\Bigg|_{{\bf r}_{\perp i} \to {\bf r}_\perp}
+(k_2 \leftrightarrow k_4) \nonumber \\
&&=
\frac{2\pi T N(0)}{\sqrt{2}} \delta_{k_1+k_3,k_2+k_4}
\int \prod_{i=1}^3 d\rho_i 
f(\sum_{i=1}^3 \rho )
\left.   \Big \langle 4 ({\rm Re} \zeta^2)^4 I_{4} (\{\lambda_i \}) 
\right|_{\lambda_4=0, \lambda_{i\ne4}= \rho_i \zeta^*/\tau_H }
   \Big \rangle_{\bf \hat{p}} + c.c., 
\end{eqnarray}
\begin{eqnarray} 
&&\left. 
\int \frac{d^2 r_{\perp}}{S_H} \hat{K}_{4bc}( \{ {\bf \Pi}_i \}) 
u_{0k_1}^*({\bf r}_{\perp 1})u_{0k_2}({\bf r}_{\perp 2})
u_{0k_3}^*({\bf r}_{\perp 3})u_{0k_4}({\bf r}_{\perp 4})
\right|_{{\bf r}_{\perp i} \to {\bf r}_\perp} \hspace{6.5cm} \nonumber \\
&&=
-\frac{2\pi T N(0)}{\tau \sqrt{2}} \int \prod_{i=1}^4 d\rho_i 
f(\sum_{i=1}^4 \rho )
\int \frac{d^2 r_\perp}{S_H}
\left< 
|w_{\bf p}|^2
e^{{\rm i}(\rho_1  {\bf v \cdot \Pi}_1^* + \rho_2 {\bf v \cdot \Pi}_2 )} 
u_{0k_1}^*({\bf r}_{\perp 1})u_{0k_2}({\bf r}_{\perp 2})
\right>_{\bf \hat{p}} \nonumber \\
&&\hspace{3cm}\times
\left.
\left< 
|w_{\bf p}|^2 e^{{\rm i}(\rho_3 {\bf v' \cdot \Pi}_3^* 
+\rho_4 {\bf v' \cdot \Pi}_4 )} 
u_{0k_3}^*({\bf r}_{\perp 3})u_{0k_4}({\bf r}_{\perp 4})
\right>_{\bf \hat{p}'}
\right|_{{\bf r}_{\perp i} \to {\bf r}_\perp} +(k_2 \leftrightarrow k_4) \nonumber \\
&&=
-\frac{2\pi T N(0)}{\tau \sqrt{2}} \delta_{k_1+k_3,k_2+k_4}  \int \prod_{i=1}^4 d\rho_i 
f(\sum_{i=1}^4 \rho )
\left.   \Big \langle 4 ({\rm Re} \zeta^2)^2 ({\rm Re} \xi^2)^2 
I_{4} (\{\lambda_i \}) 
\right|_{\lambda_{1,2}=\rho_{1,2} \zeta^*/\tau_H ; \, 
\lambda_{3,4}=\rho_{3,4} \xi^*/\tau_H}
   \Big \rangle_{\bf \hat{p},\hat{p}'} + c.c., \nonumber \\
\end{eqnarray}
where the suffix $\lambda_{i,j}=\rho_{i,j}\zeta^*/\tau_H$ implies 
$\lambda_i=\rho_i \zeta^*/\tau_H$ and $\lambda_j=\rho_j \zeta^*/\tau_H$.
Here the function $I_4$ for the quartic term is given by 
\begin{equation}
\ln \Big( I_4(\{ \lambda_i \}) \Big)= 
-\frac{1}{4}\sum_{i=1}^4 |\lambda_i|^2
-\frac{1}{8}(\lambda_{13}^{*2}+\lambda_{24}^2)
-\frac{1}{4}(\lambda_1^*+\lambda_3^*)(\lambda_2+\lambda_4)
-\frac{r_H}{2}(k_{13}\lambda_{13}^*-k_{24}\lambda_{24}), 
\end{equation}
where $\xi={\hat p}'_x+{\rm i}{\hat p}'_y$, $k_{ij}=k_i-k_j$, 
and $\lambda_{ij}=\lambda_i-\lambda_j$. 
Finally we have a quartic term 
\begin{eqnarray}
{\cal F}_4 &=& 
\frac{N(0)}{2 \sqrt{2} {S_H}} 
 \sum_j \sum_{\{k_i\}} \delta_{k_1+k_3,k_2+k_4} 
 V_4(\{ k_i \}) e^{-\frac{r_H^2}{4}(k_{13}^2+k_{24}^2)} 
 \phi^*_{0, k_1,j} \phi_{0, k_2,j} \phi^*_{0, k_3,j} \phi_{0, k_4,j}, \nonumber \\
&&
\label{F43}
\end{eqnarray}
\begin{eqnarray}
 V_{4}( \{ k_i \} ) &=& 
2\pi T 
     \int \prod_{i=1}^{3} d\rho_i 
        f \Big( { \textstyle \sum_{i=1}^3 \rho_i } \Big) \cdot
\left.   \Big \langle 4 ({\rm Re} \zeta^2)^4 I_{4} (\{\lambda_i \}) 
\right|_{\lambda_4=0; \, \lambda_i=\rho_i \zeta^*/\tau_H }
   \Big \rangle_{\bf \hat{p}} \nonumber \\
&& 
 -\frac{2\pi T}{\tau} 
     \int \prod_{i=1}^{4} d\rho_i   
    f \Big( { \textstyle \sum_{i=1}^4 \rho_i } \Big) \cdot
\left.   \Big \langle 4 ({\rm Re} \zeta^2)^2 ({\rm Re} \xi^2)^2 I_{4} (\{\lambda_i \}) 
\right|_{\lambda_{1,2}=\rho_{1,2} \zeta^*/\tau_H; \, 
\lambda_{3,4}=\rho_{3,4} \xi^*/\tau_H }
   \Big \rangle_{\bf \hat{p},\hat{p}'} 
 + c.c.. 
\label{V41} 
\end{eqnarray}
We will show later (see Fig.5 (a)) that, consistently with the neglect of 
the second term of eq.(\ref{Dd}), the second contribution to eq.(\ref{V41}) 
arising from ${\hat K}_{\rm {4bc}}$ is safely negligible compared with 
the first ${\hat K}_{\rm {4a}}$ term in the relatively clean case with 
$2 \pi T_{c0} \tau > 10$ of our interest.

\subsection{C. 6th order term}
\begin{figure}[t]
\scalebox{0.6}[0.6]{\includegraphics{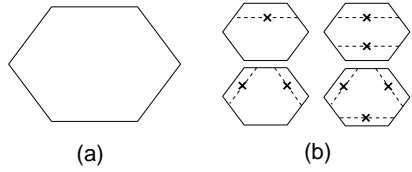}}
\caption{  Diagrams expressing the 6th order term of the GL functional. } 
\label{fig:6th}
\end{figure}

\begin{figure}[t]
\scalebox{0.6}[0.6]{\includegraphics{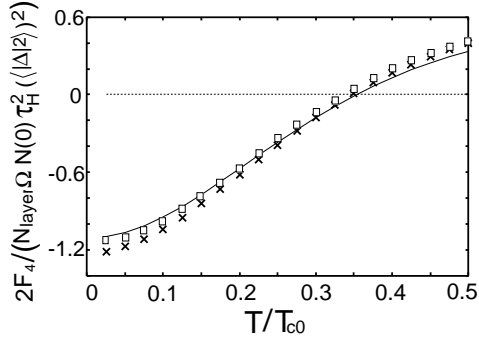}}
\caption{ Results on ${\tilde F}_4 \equiv 2{\cal F}_4/[N_{\rm {layer}} \Omega N(0) \tau_H^2 (\langle|\Delta^{(0)}|^2\rangle)^2 ]$ calculated under three different conditions and in the pure $(\tau T_{c0}=\infty)$ limit.
In the open-square symbols, the square vortex lattice is assumed with the 
nonlocal contribution in ${\tilde F}_4$ included, 
while the triangular lattice is assumed in both the crossed symbols with the 
nonlocal contribution and the solid curve with no nonlocal one. 
Note that $T^*$ at which ${\tilde F}_4 =0$ is insensitive to the 
nonlocal contribution and to the types of lattices.} 
\label{fig:V4_nonloc}
\end{figure}
When we restrict the pair-field into the LLL subspace, the 
6th order term of the GL functional are expressed as follows 
\begin{eqnarray}
{\cal F}_6 &=& 
\left. -\frac{1}{3}\sum_j \int d^2 r_{\perp}
\hat{K}_6( \{ {\bf \Pi}_i \})
(\Delta_j^{(0)}({\bf r}_{\perp 1}) \Delta_j^{(0)}({\bf r}_{\perp 3}) 
\Delta_j^{(0)}({\bf r}_{\perp 5}))^* 
\Delta_j^{(0)}({\bf r}_{\perp 2}) 
\Delta_j^{(0)}({\bf r}_{\perp 4}) 
\Delta_j^{(0)}({\bf r}_{\perp 6})
\right|_{{\bf r}_i \to {\bf r}} \\
\label{F61}
&=&
-\frac{1}{3S_H^2} \sum_{j,\{k_i\}} 
\phi^*_{0,k_1,j} \phi_{0,k_2,j} \phi^*_{0,k_3,j} 
\phi_{0,k_4,j} \phi^*_{0,k_5,j} \phi_{0,k_6,j} \nonumber \\
&&\hspace{3cm}\times \left.
\int \frac{d^2 r_{\perp}}{S_H} \hat{K}_6( \{ {\bf \Pi}_i \}) 
u_{0k_1}^*({\bf r}_{\perp 1})u_{0k_2}({\bf r}_{\perp 2})
u_{0k_3}^*({\bf r}_{\perp 3})u_{0k_4}({\bf r}_{\perp 4})
u_{0k_5}^*({\bf r}_{\perp 3})u_{0k_6}({\bf r}_{\perp 4})
\right|_{{\bf r}_{\perp i} \to {\bf r}_\perp}. \nonumber \\
\label{F62}
\end{eqnarray}
In contrast to the quartic term, 
the kernel $\hat{K}_6$ also includes diagrams (see Fig.\ref{fig:6th} (b)) 
with two or three 
impurity lines in addition to those with a single impurity line such as 
Fig.\ref{fig:quart} (b). 
Fortunately, according to the statement following eq.(\ref{V41}), 
all terms other than Fig.\ref{fig:6th} (a) may be neglected 
in the range of purity parameter 
we focus on. The diagram Fig.\ref{fig:6th} (a) is expressed as 
\begin{eqnarray}
\hat{K}_6 &=&	
2\pi T N(0) \sum_{\varepsilon}
{\rm i}s_{\varepsilon} 
\left<
\left( |w_{\bf p}|^6 \, \prod_{i=1}^{6} \frac{1}{z_i} \right)
 \Big( z_6+z_4+z_2 + \frac{z_3 z_6}{z_2-z_3+z_4} + 
\frac{z_1 z_4}{z_3-z_4+z_5} + \frac{z_2 z_5}{z_1-z_2+z_3} \Big) 
\right>_{\sigma, {\bf \hat{p}}}  \nonumber \\
&\equiv& \hat{K}_{6a} +  \hat{K}_{6b} + \hat{K}_{6c} 
+ \hat{K}_{6d} + \hat{K}_{6e} + \hat{K}_{6f} 
\label{K61}
\end{eqnarray}
where $z_i \equiv 2{\rm i}\tilde{\varepsilon}_\sigma + {\bf v \cdot \Pi}_i$ 
for 
even $i$ and $2{\rm i}\tilde{\varepsilon}_\sigma + {\bf v \cdot \Pi}_i^*$ for odd $i$.
It is easily seen that, due to the symmetry with respect to ${\bf \Pi}_i$ 
and ${\bf \Pi}_i^*$, the above expression can be represented in terms only of, 
e.g., its first and fourth terms.

Firstly, let us calculate the contribution to the GL functional of $\hat{K}_{6a}$. 
Using the parametric representation (see Appendix A), it is written as 
\begin{eqnarray} 
&&\left. 
\int \frac{d^2 r_{\perp}}{S_H} \hat{K}_{6a}( \{ {\bf \Pi}_i \}) 
u_{0k_1}^*({\bf r}_{\perp 1})u_{0k_2}({\bf r}_{\perp 2})
u_{0k_3}^*({\bf r}_{\perp 3})u_{0k_4}({\bf r}_{\perp 4})
u_{0k_5}^*({\bf r}_{\perp 3})u_{0k_6}({\bf r}_{\perp 4})
\right|_{{\bf r}_{\perp i} \to {\bf r}_\perp}\\
&&=
2\pi T N(0) \int \prod_{i=1}^5 d\rho_i 
f(\sum_{i=1}^5 \rho ) \int \frac{d^2 r_{\perp}}{S_H} \nonumber \\
&&\times\left. 
\left< |w_{\bf p}|^6 
e^{{\rm i} (\sum_{i=1,3,5}\rho_i {\bf v \cdot \Pi}_i^*
+\sum_{i=2,4}\rho_i {\bf v \cdot \Pi}_i)
}
u_{0,k_1}^*({\bf r}_{\perp1})
u_{0,k_2}({\bf r}_{\perp2})
u_{0,k_3}^*({\bf r}_{\perp3})
u_{0,k_4}({\bf r}_{\perp4}) 
u_{0,k_5}^*({\bf r}_{\perp5})
u_{0,k_6}({\bf r}_{\perp6}) 
\right>_{\bf \hat{p}}
\right|_{{\bf r}_{\perp i} \to {\bf r}_{\perp}}\\
&&=
\frac{2\pi T N(0)}{\sqrt{3}} \delta_{k_1+k_3+k_5,k_2+k_4+k_6} \int \prod_{i=1}^5 d\rho_i 
f(\sum_{i=1}^5 \rho )
\left.   \Big \langle 8 ({\rm Re} \zeta^2)^6 \, I_{6} (\{\lambda_i \}) 
\right|_{\lambda_6=0, \lambda_{i \ne 6}=\rho_{i} \zeta^*/\tau_H}
   \Big \rangle_{\bf \hat{p}}, 
\end{eqnarray}
where the function $I_6$ is given by 
\begin{eqnarray}
\ln \Big(I_6(\{ \lambda_i \}) \Big)
&=& 
-\frac{1}{4}\sum_{i=1}^6 |\lambda_i|^2 
+\frac{1}{4}(\sum_{i:{\rm odd}} \lambda_i^{*} 
+ \sum_{i:{\rm even}} \lambda_{i}) 
-\frac{1}{12}(\sum_{i:{\rm odd}} \lambda_i^{*} 
+\sum_{i:{\rm even}} \lambda_{i})^2 \nonumber \\
&&-\frac{1}{6}(\sum_{(i,j):{\rm odd}} \lambda_{ij}^{*2}
+\sum_{(i,j):{\rm even}} \lambda_{ij}^2)
+\frac{r_H}{3}(\sum_{(i,j):{\rm odd}} k_{ij}\lambda_{ij}^{*}
-\sum_{(i,j):{\rm even}} k_{ij}\lambda_{ij} ).
\end{eqnarray}

Next we examine $\hat{K}_{6d}$. 
Using the parameter representation, this term is expressed by changing the 
integral variables in the above expression as 
$\rho_2 \to \rho_2+\rho_3$, $\rho_3 \to -\rho_3$, and $\rho_4 
\to \rho_3+\rho_4$.

In this way, 
we can write the 6th order term of the GL free energy functional in the form
\begin{eqnarray}
{\cal F}_{6}
&=&
\frac{N(0)}{3 \sqrt{3} {S_H^2}} \sum_j 
\sum_{\{k_i\}} \delta_{k_1+k_3+k_5, k_2+k_4+k_6} 
V_{6}(\{k_i\}) \; e^{ -\frac{r_H^2}{6} \sum_{(i,m)} k_{im}^2 } \;
 \phi^*_{0,k_1,j} \phi^*_{0,k_3,j} \phi^*_{0,k_5,j} 
 \phi_{0,k_2,j} \phi_{0,k_4,j} \phi_{0,k_6,j}, 
 \label{F63}
\end{eqnarray}
where the summation $\sum_{(i,m)}$ is taken over the pairs 
$\{(1,3),(3,5),(5,1),(2,4),(4,6),(6,2) \}$, and  
$V_{6}$ is given by 
\begin{eqnarray}
V_{6}(\{k_i\}) &=&
-3\times2 \pi T \int \prod_{i=1}^5 d\rho_i f(\sum_{i=1}^5 \rho_i)
\Bigg\{ 
\left< 8 ({\rm Re} \zeta^2)^6 \,
\left.I_6(\{ \lambda_i \} )
\right|_{\lambda_6=0; \,  \lambda_{i\neq6}=\rho_i \zeta^*/\tau_H} 
\right>_{\bf \hat{p}} \hspace{3cm} \nonumber \\
&&\hspace{3cm}+
\left< 8 ({\rm Re} \zeta^2)^6 \,
\left.I_6(\{ \lambda_i \} )\right|
_{\lambda_6=0;\, \lambda_{1,5}=\rho_{1,5} \zeta^*/\tau_H; \, 
\lambda_{3}=-\rho_3 \zeta^*/\tau_H; \,
\lambda_{2,4}=(\rho_{2,4}+\rho_3) \zeta^*/\tau_H
} 
\right>_{\bf \hat{p}} \Bigg\}.
\label{V61}
\end{eqnarray} 

\begin{figure}[t]
\scalebox{0.6}[0.6]{\includegraphics{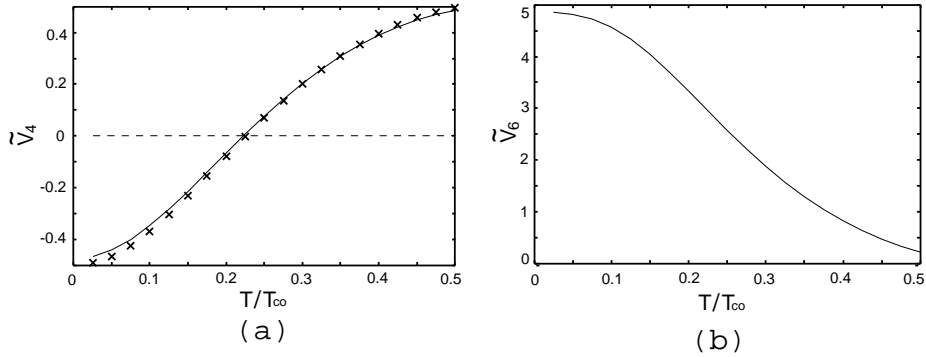}}
\caption{ Numerical results of the dimensionless coefficients 
(a) ${\tilde V}_4 = V_4(\{ k_j=0 \})/\tau_H^2$ and 
(b) ${\tilde V}_6 = V_6(\{ k_j=0 \})/\tau_H^4$ 
on the $H_{c2}$-curve at lower temperatures $T \leq 0.5 T_{c0}$. 
The value 
$(2 \pi T_{c0} \tau)^{-1}=0.05$ was commonly used. 
In (a), the cross symbols represent the result due only to 
Fig.\ref{fig:quart}(a). } \label{fig:K4K6}
\end{figure}

\begin{figure}[b]
\scalebox{0.6}[0.5]{\includegraphics{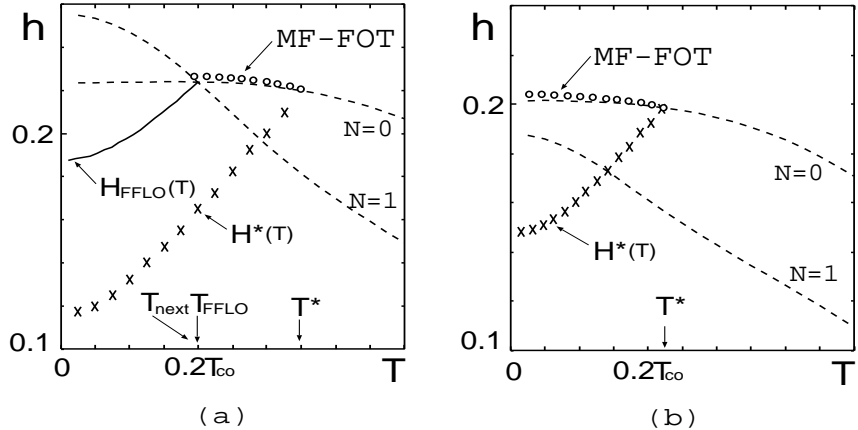}}
\caption{ High $H$ and low $T$ region of MF $H$-$T$ phase diagrams in (a) clean limit ($(2 \pi T_{c0} \tau)^{-1}=0$) and 
(b) the moderately clean case ($(2 \pi T_{c0} \tau)^{-1}=0.05$). 
See the text for further details.}
\label{fig:Hc2}
\end{figure}

\end{widetext}

In deriving the MF phase diagram and its impurity dependence, 
we will use an additional approximation below. 
Since a nonzero magnetic field plays the roles of cutting off the low $T$ 
divergences of coefficients of the higher order GL terms, 
the orbital depairing effect arising from the gauge invariant gradients 
${\bf \Pi}_j$ has been incorporated nonperturbatively in the above expressions. 
Instead, algebraic $k_{ij}$ dependences have arisen in the vertices 
$V_m(\{k_{ij}\})$ with $m=4$ and $6$. On the other hand, the Gaussian 
$k_{ij}$-dependences in ${\cal F}_4$ and ${\cal F}_6$ are direct consequences of 
restricting the pair-field into the LLL subspace and also appear 
in the familiar GL expression with spatially nonlocal higher order terms. 
That is, the additional $k_{ij}$ dependences in $V_m(\{k_{ij}\})$ can be seen 
as spatially {\it nonlocal} contributions to the higher order GL terms and affect 
the structure of vortex solid. Actually, in LLL and the case with 
no paramagnetic depairing, this nonlocality in the quartic GL term results 
in the structural transition between the rhombic and square vortex 
lattices \cite{Asahi}. 
However, an energy difference affecting the lattice structure is 
extremely small reflecting a weak structure dependence of the Abrikosov factors 
(denoted as $\beta_A$ and $\gamma_A$ below). 
For instance, as shown in Fig.\ref{fig:V4_nonloc}, 
the nonlocal correction to 
${\tilde F}_4$ and thus, to $T^*$ is negligibly small. 
Therefore, at least as far as the SC transition in the MF approximation 
at $H_{c2}$ is concerned, such nonlocal corrections are safely negligible. 
For this reason, the local approximation for the higher order terms will be 
used 
hereafter, and $V_m(\{k_{ij}\})$ will be replaced by $V_m=V_m(\{k_{ij}=0\})$.
Then, the GL functional derived microscopically takes the form 
\begin{eqnarray}
{\cal F}_{\rm loc} &=& 
N(0) \int d^2r_\perp \, \Bigg[ \sum_{q_z} a_0(q_z^2) 
|{\tilde \Delta}_{q_z}^{(0)}({\bf r}_\perp)|^2 \nonumber \\
&&\hspace{1cm} + \sum_j \Big( \frac{V_4}{2} |\Delta_j^{(0)}|^4 + 
\frac{V_6}{3}|\Delta_j^{(0)}|^6 \Big) \Bigg]. 
\label{Floc}
\end{eqnarray}

Temperature variations of the coefficients $V_4$ and $V_6$ calculated along 
the $H_{c2}(T)$-curve are shown, respectively, in Fig.\ref{fig:K4K6} (a) and (b). 
To clarify that the contributions of Fig.\ref{fig:quart} (b) and (c) are 
safely negligible, $V_4$ in Fig.\ref{fig:K4K6} (a) was calculated in terms of 
$(2 \pi T_{c0} \tau)^{-1}=0.05$, the value used in Fig.\ref{fig:Hc2} (b) below. 
The coefficient $V_4$ is negative at lower temperatures, while $V_6$ is positive 
over a broad region so that the GL expression (\ref{Floc}) with a MF-FOT 
at $H_{c2}(T)$ is well-defined.

\subsection{D. Mean field Phase diagram}

Below, the MF phase diagram will be examined based on the functional, 
eq.(\ref{Floc}). 
First, let us neglect a possibility of an FFLO-like state and assume 
a straight vortex solid independent of $j$ as the MF solution. 
Then, the first term in the bracket of eq.(\ref{Floc}) is replaced by 
$\sum_j a_0(0) |\Delta_j|^2$, and the MF solution is obtained in a standard way. 
The character of the MF transition at $H_{c2}(T)$-line changes with increasing 
field from the second order one to a discontinuous one at a temperature $T^*$ 
where $V_4$ becomes negative (see Fig.\ref{fig:V4_nonloc} and 
Fig.\ref{fig:K4K6} (a)), reflecting that the 
spin depairing is more effective upon cooling and with increasing $H$. 
To obtain MF results in $T < T^*$, higher order terms are necessary in the GL 
expression. According to Fig.\ref{fig:K4K6} (b), the coefficient $V_6$ 
is positive over a 
broad temperature range, and thus, the expression eq.(\ref{Floc}) terminated 
at the 6th order term will be sufficient for our purpose. 
Further, let us introduce the effective coefficients ${\tilde b} = V_4 \beta_A$ 
of the quartic term and ${\tilde c}=V_6 \gamma_A$ of the 6th order term, respectively, 
where 
\begin{eqnarray}
\beta_A &=& 
\frac{\left< |\Delta_j^{(0)}({\bf r}_\perp)|^4 \right>}
{( \left<|\Delta_j^{(0)}|^2 \right>)^2}, \nonumber \\
\gamma_A &=& 
\frac{\left< |\Delta_j^{(0)}({\bf r}_\perp)|^6 \right>}
{( \left<|\Delta_j^{(0)}|^2 \right>)^3}. 
\end{eqnarray}

\begin{figure}[t]
\scalebox{0.7}[0.7]{\includegraphics{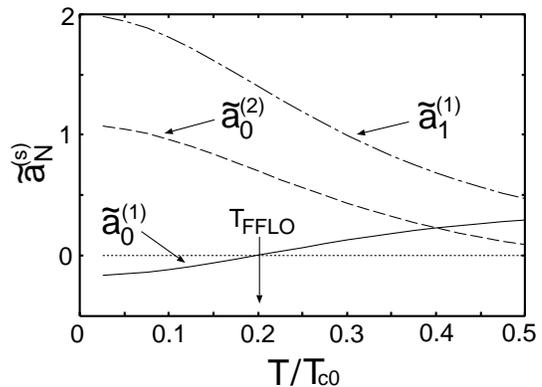}}
\caption{ Temperature dependences of $\tilde{a}_0^{(1)}$ and $\tilde{a}_0^{(2)}$ 
on $H_{c2}(T)$ and near $T_{\rm FFLO}$, where 
$\tilde{a}_0^{(s)}=a_0^{(s)}/\tau_H^{2s}$ $(s=1,2)$, 
and the pure limit was assumed. 
The corresponding curve of $\tilde{a}_1^{(1)}=a_1^{(1)}/\tau_H^2$ 
is also shown for a convenience.} \label{fig:aN}
\end{figure}

Then, the following MF results in $T < T^*$ are found. 
First, the MF transition point in $T < T^*$ and in LLL is determined by the condition
\begin{equation}
 a_0(0)= \frac{3 {\tilde b}^2}{16 {\tilde c}} \equiv a_{0, c}, 
 \label{a0c}
\end{equation}
while the supercooling (superheating) point is given by $a_0(0)=0$ 
($a_0(0)={\tilde b}^2/(4 {\tilde c})$). 
Next, the energy barrier $U_{\rm barr}$ between the $|\Delta^{(0)}|=0$ solution 
and the jump value of $|\Delta^{(0)}|$ at the transition, 
$|\Delta^{(0)}|_c= \sqrt{3 |{\tilde b}|/4{\tilde c}}$, is given by 
\begin{equation} 
U_{\rm barr} = N(0) \frac{|{\tilde b}|^3}{48 {\tilde c}^2}. 
\end{equation}
Further, by calculating the mean squared amplitude 
$\left< |\delta \Delta^{(0)}|^2 \right>$ 
of the Gaussian fluctuation $\delta \Delta^{(0)}$ when $a_0(0)=a_{0,c}$ 
and in 2D limit, one also finds 
\begin{equation}
\left< |\delta \Delta^{(0)}|^2 \right> 
\propto \frac{T_{c2} \, |\Delta^{(0)}|_c^2}{U_{\rm barr}}, 
\label{flu}
\end{equation}
where $T_{c2}$ is the MF transition temperature. 
Thus, the fluctuation strength is enhanced with decreasing $|{\tilde b}|$ 
and increasing ${\tilde c}$ and, as expected, is measured at $T_{c2}$ by the 
inverse of the energy barrier. Hence, if this MF-FOT occurs as a true 
FOT in real systems, a clear hysteresis is expected in a system with weaker fluctuation. 

However, in higher fields and lower temperatures where the spin depairing 
becomes more important, an FFLO-like helical vortex solid within LLL may 
become more favorable. As far as the width $a_{0,c}$ is sufficiently small, 
the $q_z$-dependent terms have only to be incorporated in the quadratic terms in 
$\Delta_j^{(0)}$. That is, this {\it structural} transition line $H_{\rm FFLO}(T)$ 
between the FFLO-like solid and the straight vortex solid may be discussed 
within the coefficient $a_0(q_z^2)$. Actually, according to the calculation 
results of $V_4$ and $V_6$ in Fig.\ref{fig:K4K6}, $a_{0,c}$ in Fig.\ref{fig:Hc2} is 
at most of the order of $10^{-2}$. Assuming a helical state with the $j$-dependence 
$\Delta_j^{(0)} \sim e^{{\rm i}q_z js}$ in $H > H_{\rm FFLO}$, 
a second order structural transition line $H_{\rm FFLO}(T)$ is obtained 
within LLL according to eq.(\ref{aN2}) as the curve $a_0^{(1)}(T)=0$ if 
$a_0^{(2)}(T) > 0$ there. Numerical results on $a_0^{(1)}$ and $a_0^{(2)}$ 
are shown in Fig.\ref{fig:aN}.
We find that $a_0^{(2)}$ near $H_{c2}$ is always positive 
along the $H_{\rm FFLO}(T)$-line and increases upon cooling at a fixed field 
above $H_{\rm FFLO}(T)$. Further, since the paramagnetic depairing effect is 
enhanced with increasing field and decreasing temperature, as the example in 
Fig.\ref{fig:Hc2} shows, a possible $H_{\rm FFLO}(T)$-curve should decrease 
upon cooling.

 Now, the MF phase diagrams Fig.\ref{fig:Hc2} 
following from the GL coefficients derived above and 
for two different impurity strengths ($(\tau T_{c0})^{-1}$-values) 
will be explained. 
The used value of paramagnetic parameter $\mu_0 H_{c2}^{\rm orb}(0)/2 \pi T_{c0}$, corresponding to the Maki parameter, is 0.8, which leads to the value $T^*=0.36 T_{c0}$ in the pure limit (Fig.6 (a)), 
where $H_{c2}^{\rm orb}(0)=0.56 \phi_0/(2 \pi \xi_0^2)$ is the 2D orbital 
limiting field at $T=0$ in pure limit, and $\xi_0=k_F/2\pi m T_{c0}$ 
is the coherence length. The field values in the figures were normalized by 
$H_{c2}^{\rm orb}(0)$ (i.e., $h=H/H_{c2}^{\rm orb}(0)$). 
The curve $H^*(T)$ indicated by the cross symbols is defined 
by the condition $V_4=0$ and may not be directly seen in experiments. 
In contrast, the portion (open circles) in $T < T^*$ of $H_{c2}(T)$ on which 
the MF SC transition is discontinuous is experimentally measurable together 
with (if any) the second order transition line $H_{\rm FFLO}(T)$ (solid curve) 
to the FFLO vortex solid. In the temperature regions where the MF-FOT does 
not occur, 
the higher one of the dashed curves indicated by $N=0$ or $N=1$ becomes 
the $H_{c2}(T)$-line, on which $a_0(0)$ or $a_1(0)=0$, 
and the MF transition there is of second order. 

It will be important to, 
in relation to real phase diagrams of related materials, 
understand how the $H_{\rm FFLO}(T)$ curve and the characteristic temperatures 
are affected by the impurity strength. By comparing Figs.6 (a) and (b) 
with each other, 
the region $H_{\rm FFLO}(T) < H < H_{c2}(T)$ is found to be easily lost by a 
slight increase of impurity strength $(\tau T_{c0})^{-1}$. 
In contrast, the onset $T^*$ of the MF-FOT behavior is relatively insensitive 
to the sample purity. Nevertheless, when $(2 \pi T_{c0} \tau)^{-1}$ goes 
beyond $0.095$ while the value $\mu_0 H_{c2}^{\rm orb}(0)/2 \pi T_{c0} = 0.8$ 
was kept, 
the MF-FOT region at $H_{c2}(T)$ is also lost, and the MF transition at $H_{c2}(T)$ 
is continuous at all temperatures. This result is contrast 
to other works \cite{Agterberg,Maki} in which the presence of a MF-FOT was argued 
under the use of dirty limit. We find that, instead, the FOT obtained 
in the dirty limit \cite{Maki} never occurs in $T \to 0$ limit when $E_F \tau > 1$ 
under which the usual dirty limit may be valid. On the other hand, 
the results in ref.\cite{Agterberg} are derived by completely neglecting 
the orbital depairing and are not comparable with the present ones. 
Further, we stress that, in contrast to results given in a previous 
work \cite{Houzet} taking account of both the orbital and spin depairing effects, 
the results in Fig.\ref{fig:Hc2} imply that always $T^* > T_{\rm FFLO}$ 
in the present quasi 2D case under a perpendicular magnetic 
field. 

Since, as already mentioned, the width $a_{0,c}$ in Fig.\ref{fig:Hc2} is 
relatively small, the MF-FOT there may be regarded as being weak. 
However, it does not mean a strong fluctuation. 
Actually, in systems with a large $N(0)$ in zero field such as CeCoIn$_5$, 
the fluctuation strength $T/U_{\rm {barr}}$ itself becomes extremely small 
in the low $T$ region of our interest. On the other hand, the magnetization 
jump value $\Delta M_c$ at the MF-FOT should be quite small compared 
with the applied field in order to justify our neglect of 
a spatially varying {\it internal} magnetic field.
In CeCoIn$_5$ under 
an applied field in the tesla range, this condition is well satisfied \cite{Tayama} 
(see also Sec.IV).

In $T < T_{\rm next}$ where $a_1(0) < a_0(0)$ ($< a_2(0)$), 
the $H_{c2}(T)$ line and hence, 
the vortex lattice itself just below it are determined by the next 
lowest ($N=1$) LL. According to the $a_1^{(1)}(T)$-curve in Fig.\ref{fig:aN}, 
this $N=1$ state is not a modulated FFLO-like state but a straight vortex solid. 
Further, note that $T_{\rm next}$ and $T_{\rm FFLO}$ are close to each other 
(see Fig.\ref{fig:Hc2} (a)). These results imply that, in quasi 2D materials 
under fields perpendicular to the SC layers, the presence itself of the FFLO-like 
vortex solid is very 
subtle even in pure limit. Thus, a competition between the FFLO-like solid within LLL 
and a solid within the $N=1$ LL has to be examined just above 
the $H_{\rm FFLO}(T)$ line. Since this is an issue of a transition between 
vortex lattice structures defined within the planes perpendicular to the field, 
a detailed description of the stable vortex lattices in $d$-wave pairing cases is 
required to address this. As already mentioned, however, the nonlocality, 
at least, of the sixth order GL term affecting the in-place lattice structure 
was neglected in this paper. 
We will postpone a study of structural transitions to higher LL 
solids in $H > H_{\rm FFLO}$ to our future study. 

\subsection{E. Properties of Helical Vortex Solid}

Here, we briefly comment on linear responses and elastic properties 
in vortex solid phases, primarily in an FFLO-like helical solid with 
a modulation $\Delta_j \simeq {\tilde \Delta}_{q_m} e^{{\rm i}q_m js}$ in LLL 
with $q_m \neq 0$. 
To examine the electro-magnetic linear responses in an ordered vortex lattice phase, 
we have only to focus on the gradient terms with an {\it external} gauge 
fluctuation substituted and to examine the Gaussian fluctuation around a MF 
solution of $\Delta_j$. An appropriate form, consistent with the above microscopic 
analysis, of the gradient energy will be
\begin{eqnarray}
{\cal F}_{\rm grad} &=& 
\int d^3r \, \Delta^*({\bf r}_\perp,z) \Bigg[ {\cal A}({\bf \Pi}^2) 
\Bigg(-{\rm i}{\partial \over {\partial z}}-a_z \Bigg)^4 \;
\nonumber \\ 
&-& 2 {\cal B}({\bf \Pi}^2) \Bigg(-{\rm i}{\partial \over {\partial z}}-a_z \Bigg)^2 \;
\nonumber \\ 
&+&  ({\cal A}({\bf \Pi}^2))^{-1} ({\cal B}({\bf \Pi}^2))^2 \Bigg] 
\Delta({\bf r}_\perp, z),
\label{Fgrad1}
\end{eqnarray} 
where a continuous variable $z$ was used for the coordinate parallel to the field. 
If considering the SC response in the $x$ or $y$ direction, ${\bf \Pi}$ needs 
to be accompanied by a gauge fluctuation ${\bf a}={\bf a}_\perp+a_z {\bf \hat{z}}$
in the form ${\bf \Pi} - {\bf a}_\perp$. 
Within the Gaussian approximation for the fluctuation, no cross terms like 
${\bf a}_\perp \, a_z$ appear because any term $\propto {\bf a}_\perp$ 
becomes off-diagonal with respect to the LLs and hence, zero after spatial averaging. 
Hence, the linear responses in the parallel and perpendicular directions can 
be considered independently. 

First, let us consider the response parallel to the field in which ${\bf a}_\perp=0$. 
If the spatial variations perpendicular to ${\bf H}$ of the MF pair-field solution 
are described within LLL, the argument ${\bf \Pi}^2$ in ${\cal A}$ and 
${\cal B}$ can be replaced by $r_H^{-2}$. 
For the moment, let us focus on the helical solid phase in which 
${\cal B}(r_H^{-2}) > 0$. Assuming the fluctuation of $\Delta_j$ to be 
dominated by that of its phase $\phi$, the fluctuation part of 
$\delta {\cal F}_{\rm grad}$ simply becomes 
\begin{eqnarray}
\delta {\cal F}_{\rm grad} &=& 
{\cal A}(r_H^{-2}) \langle|\Delta|^2 \rangle \! \int d^3r 
\Bigg[ \Bigg(\!(\partial_z \phi)^2 - q_m^2 \;
\nonumber \\ 
 && \hspace{3cm} -2 \partial_z \phi a_z \Bigg)^2 + (\partial_z^2 \phi)^2 \Bigg] \;
\nonumber  \\ 
&\simeq& 4 {\cal B}(r_H^{-2}) \langle |\Delta|^2 \rangle 
\int d^3r (\partial_z \delta \phi 
- a_z)^2. 
\label{Fgrad2}
\end{eqnarray}
Here, $\delta \phi$ is the phase fluctuation, and higher gradient terms were 
omitted in the expression following the second equality. 
Further, the fact that the MF value $q_m^2$ of $q_z^2$ is 
${\cal B}(r_H^{-2})/{\cal A}(r_H^{-2})$ was used. 
As seen in the seminal work \cite{LO}, the {\it same} form as above occurs 
in the case with no orbital depairing in an isotropic 3D system (see eq.(31) 
in ref.\cite{LO}). 
This familiar result (\ref{Fgrad2}) implies that, in the helical solid state, 
there is no {\it stationary} current even along ${\bf H}$ and that a supercurrent 
$\propto a_z$ can flow in this direction, although the superfluid rigidity 
$\propto {\cal B}(r_H^{-2})$ vanishes with approaching $H_{\rm FFLO}(T)$. 

The absence of SC response in $x$ and $y$ directions can be seen, 
as well as in the ordinary ($q_m=0$) case, most easily within the phase-only 
approximation \cite{RI95}. Noting that any term accompanied by the in-plane 
periodic variation in a vortex lattice disappears after averaging spatially 
and that the {\it uniform} fluctuation part of $\partial_\perp \delta \phi$ 
is given by $r_H^{-2} ({\hat z} \times {\bf s})$, 
one finds that the fluctuation of ${\bf \Pi}^2$ 
in the limit of vanishing wave number can be expressed {\it everywhere} 
in ${\cal A}$ and ${\cal B}$ as 
$r_H^{-4} (({\hat z} \times {\bf s}) - {\bf a}_\perp)^2$, 
where ${\bf s}$ is a uniform in-plane displacement of a vortex lattice. 
After integrating over ${\bf s}$, ${\bf a}_\perp$ disappears together 
in the resulting fluctuation free energy, implying no SC response. 
This non-SC perpendicular response 
is a consequence of the establishment of the Josephson relation 
${\bf a}_\perp=r_H^{-2}({\hat z} \times {\bf s})$ 
(leading to an electric field ${\bf E}$ proportional to a vortex velocity 
${\bf v}_\phi$, i.e., ${\bf E} = - {\bf v}_\phi \times {\bf H}$) 
and is quite different from the corresponding one in the case with 
no orbital depairing in the {\it ideal} isotropic 3D system which is 
a consequence of {\it spontaneous} formation of a direction of the 
helical modulation \cite{com2}. 

Next, let us briefly discuss the vortex elastic energy. 
The shear elastic energy for in-plane shear distortions is obtained, 
as in the case of straight vortex lattice, from the phase fluctuation energy. 
It was already shown in ref.\cite{RI90s} that the {\it quartic} dispersion 
on the 2D wave vector ${\bf q}_\perp$ of the massless phase fluctuation and hence, 
the way of identifying the shear elastic mode with the phase fluctuation are 
unaffected by any change of the higher order terms of the GL model. 
No calculation of shear modulus will not be performed here, and we simply assume, 
just for qualitative consideration, the resulting shear energy term to be 
positive definite. In contrast, some comment will be necessary on the tilt energy 
which should change through $H_{\rm FFLO}(T)$. 
For simplicity, we focus on its expression in type II limit with no internal 
gauge fluctuation \cite{Ikeda1,Moore}. 
According to eq.(\ref{Fgrad2}), the tilt energy in type II limit takes the form 
\begin{equation}
E_{\rm tilt} = {\cal A}(r_N^{-2}) 
\langle |\Delta|^2 \rangle 
\Bigg(\frac{2 \pi H}{\phi_0} \Bigg)^2 \sum_{\bf q} 
\frac{k_z^2(k_z^2 + 4q_m^2)}{q_\perp^2} 
|{\bf s}_{\bf q}^T|^2, 
\label{tilt}
\end{equation}
where $|k_z|=||q_z|-q_m|$, and the relation \cite{Ikeda1,Moore,RI90s} 
$\delta \phi_q=-i (q_\perp^{-2} r_H^{-2})({\bf q}_\perp \times {\bf s}_q^{T})_z$ 
between the transverse displacement ${\bf s}^T$ and the phase fluctuation was used. 
On the other hand, in the straight vortex solid (i.e., in $H < H_{\rm FFLO}(T)$) 
where ${\cal B}(r_H^{-2}) < 0$, the dispersion $k_z^2(k_z^2+4q_m^2)$ in 
eq.(\ref{tilt}) is replaced by $2 |{\cal B}(r_H^{-2})| q_z^2/{\cal A}(r_H^{-2})$. 
Thus, the macroscopic properties in both the vortex solids are qualitatively 
the same as each other.

However, the tilt rigidity decreases with approaching $H_{\rm FFLO}(T)$. 
In particular, the $k_z^2/q_\perp^2$ dependence of the tilt modulus just 
on $H_{\rm FFLO}(T)$ implies a {\it short}-ranged phase coherence of 
the vortex solid there which is consistent with the vanishing superfluid 
rigidity on $H_{\rm FFLO}(T)$. Because this absence of SC order in the 
{\it critical} vortex solid (i.e., the case just on $H_{\rm FFLO}(T)$) is 
due to a softening (or weakening) of an elastic modulus, an enhanced peak 
effect due to an increase of the critical current is expected near $H_{\rm FFLO}(T)$, 
as well as the ordinary one near $H_{c2}(T)$ \cite{LO2}, in real materials 
with random pinning effects. This conjecture may be useful for searching 
a transition curve to the FFLO-like vortex solid. 

\section{III. Study of genuine phase diagram} 

In this section, the real phase diagram of systems described by eq.(\ref{Floc}) 
is studied by including the fluctuation effects. We will focus here on the range 
$H^*(T) < H < H_{\rm FFLO}(T)$ and hence, rewrite the first term in the bracket of 
eq.(\ref{Floc}) into 
$\sum_j ( a_0 |\Delta_j^{(0)}|^2 + \gamma_0 |\Delta_j^{(0)}- \Delta_{j+1}^{(0)}|^2 )$ 
with $\gamma_0 > 0$. 
Although we have not extended our simulation work to (if any) the range 
$H > H_{\rm FFLO}$ defined within the lowest LL, fluctuation effects  
similar to those in $H < H_{\rm FFLO}$ should be also expected 
in such higher fields (see below). 
The partition function we should examine is 
\begin{equation}
Z={\rm Tr}_\psi \exp(-{\cal F}),
\label{statsum}
\end{equation}
where the functional ${\cal F}={\cal F}_{\rm loc}/T$ is rewritten as 
\begin{equation}
{\cal F}= \sum_j \int d^2r \Biggl( \alpha |\Psi_j|^2 + 
\gamma |\Psi_j-\Psi_{j+1}|^2 - \frac{|\beta|}{2} |\Psi_j|^4 
+ \frac{1}{3}|\Psi_j|^6 \Biggr), 
\label{Fsim}
\end{equation}
where $\Psi_j({\bf r})$ is the rescaled order parameter field defined within LLL, 
$\gamma \geq 0$, 
\begin{eqnarray}
\alpha &=& \biggl(\frac{r_H^2 N(0)}{T} \biggr)^{2/3} 
\frac{a_0}{V_6^{1/3}} \simeq \alpha_0(T) \frac{H-H_0}{H_0}, \nonumber \\
|\beta| &=& \biggl(\frac{r_H^2 N(0)}{T} \biggr)^{1/3} 
\frac{|V_4|}{V_6^{2/3}},
\label{coef}
\end{eqnarray}
and the in-plane coordinates ${\bf r}$ were normalized 
like ${\bf r}/r_H \to {\bf r}$. Further, 
$H_0$ denotes $H_{c2}(0)$ in the case with no MF-FOT. 
Since, as mentioned earlier, $a_{0,c}$ measuring the difference $(H_{c2}-H_0)/H_0$ 
is small for the Maki parameter value used in this paper, $H_0$ will not be 
distinguished from $H_{c2}$ below. Note that, except a numerical factor, 
$|\beta|^{3}$ is identical with $U_{\rm {barr}}/T$. 
The $T$-dependent parameter $|\beta|^{-1}$ measures the fluctuation strength, 
while the magnetic field dependence is assumed only in $\alpha$. 
Thus, a change of $|\beta|$ can be regarded near $H_{c2}$ as a change only 
of $T$ in eq.(\ref{statsum}) under the same ${\cal F}_{\rm loc}$ 
with {\it fixed} values of $a_0$, $\gamma_0$, $V_4$, and $V_6$, 
if the magnetic field variable $\alpha$ is appropriately rescaled, 
and a difference in the 
anisotropy (i.e., $\gamma_0$-value) plays no important role. 

\begin{figure}[t]
\scalebox{0.4}[0.5]{\includegraphics{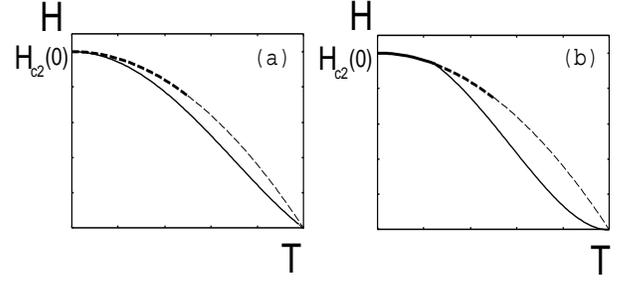}}
\caption{
Two candidates of a schematic $H$-$T$ phase diagram of bulk type II superconductors 
with paramagnetic depairing in the case with thermal fluctuation and with 
no vortex pinning effect. 
The solid curve, the thick dashed curve, and thin dashed one are the melting line 
 $H_m(T)$ of vortex solids, the MF-FOT curve on $H_{c2}(T)$, 
and the ordinary $H_{c2}$ curve of the second order MF transition, respectively. 
Dashed curves are not genuine transition lines. 
A possible presence of the structural transition to an FFLO-like vortex solid, 
neglected here for simplicity, 
does not lead to any essential change on $H_m(T)$ and $H_{c2}(T)$ 
} 
\label{fig:phaseDG}
\end{figure}

\subsection{A. Theoretical consideration}

First, let us explain how an ordinary physical picture on low energy fluctuations 
and their effect on orderings lead to the absence of a genuine transition at $H_{c2}$. 
Like in the familiar case \cite{Ikeda1,Ikeda2} with a second order MF transition 
at $H_{c2}$, let us start from a description 
deep in the ordered phase. 
First, since an inclusion of the orbital depairing requires the presence of 
field-induced vortices, a low energy excitation in the ordered phase is 
inevitably an elastic mode of a vortex solid. As already noted in the final part of 
Sec.II, 
it is clear that the form of the elastic energy and the relation 
between the phase fluctuation and the vortex displacement \cite{Moore} 
are essentially unaffected by differences in the forms of higher order GL terms. 
Then, it is clear from the previous works \cite{Moore,Ikeda1} 
that the phase fluctuation is marginally relevant even in 3D 
case, and that the rigidity controlling the {\it quasi} long-ranged phase coherence 
in the vortex solid is the shear modulus.\cite{Ikeda1} 
That is, if the vortex solid is melted at $H_m$ below $H_{c2}$, 
we have only short-ranged orders for both the phase and the vortex position 
in the vortex liquid above $H_m$ with no finite shear modulus, and thus, 
the vortex liquid should be continuously connected with the normal phase 
above $H_{c2}$ \cite{Ikeda1,Ikeda2}. 
In this sense, the MF-FOT at $H_{c2}$ should not occur. 
Further, to understand this from another point of view, let us note that 
the quasi 2D SC order parameter in the lowest LL 
has the form~\cite{Tesanovic} 
\begin{equation}
\Psi(\xi, z) = {\cal A}(z) e^{-y^2/(2 r_H^2)} \prod_{i=0}^{N_s-1}
(\xi - \xi_i(z)), 
\end{equation} 
where $\xi=x+{\rm i}y$, $z=js$, $\xi_i(z)$ 
is the complex coordinate perpendicular to 
${\bf H}$ of the $i$-th vortex, and a Landau gauge was assumed for the 
external gauge field. 
Since the vortex positions are highly disordered above $H_m$, 
the fluctuation effect 
above $H_m$ is essentially described only by the amplitude ${\cal A}(z)$. 
However, ${\cal A}(z)$ depends only on $z$ {\it irrespective of} the form of 
the higher order GL terms. 
That is, since the amplitude 
fluctuation itself has a reduced dimensionality and is 1D-like in 3D 
systems \cite{RI90f} even in the present case, the amplitude fluctuation 
will push the vortex lattice freezing field $H_m$ down to a lower 
field, and the MF-FOT should not be realized as a true FOT in 
real 3D systems \cite{comdia}. 

Once noting that the $q_z^2$-form of the tilt elastic term in $H < H_{\rm FFLO}$ 
is replaced in $H > H_{\rm FFLO}$ by $(|q_z|-q_m)^2$, the above argument precluding 
a genuine FOT at $H_{c2}$ can be applied to such higher fields with no modification. 
Note also that the continuous transition at $H_{\rm FFLO}$ is a {\it structural} 
transition between the ordered vortex solids and hence, together with FFLO-like 
states themselves, will not be lost due to the SC fluctuation {\it as far as} 
$H_m > H_{\rm FFLO}$ (see Sec. IV).

The theoretical discussion given above implies that, as far as $H_m < H_{c2}$, 
a genuine transition at $H_{c2}$ cannot occur. 
Then, it is at least natural to expect that, following the familiar case 
in $T > T^*$ with a second order $H_{c2}$-transition, $H_m < H_{c2}$ at any nonzero 
$T$ and hence that, as in 
Fig.\ref{fig:phaseDG} (a), a genuine transition at $H_{c2}$ never occurs at 
{\it any} nonzero 
temperature. However, caution will be necessary in the present case 
with a MF-FOT. For instance, a $H_m$ defined in terms of the Lindemann criterion 
\cite{Moore} may lie above $H_{c2}$ in a case with weak enough fluctuation because 
any elastic modulus, proportional in the lowest LL to some power of 
$\langle |\Psi|^2 \rangle$, 
is nonvanishing on approaching $H_{c2}$ from below. Of course, there will be no 
possibility that the actual $H_m$ lies above $H_{c2}$. Then, one may consider 
another possible phase diagram Fig.\ref{fig:phaseDG} (b) realizable for a case 
with weak enough fluctuation, in which, 
reflecting a reduction of thermal fluctuation upon cooling, 
$H_m(T)$ at low enough temperatures virtually coincides with $H_{c2}$, and {\it both} 
a large jump of magnetization (reflecting a large condensation energy) and 
a {\it tiny} hysteresis, arising from the vortex lattice melting, will be seen 
at $H_{c2}$ on the solid curve. An important point is that, even in the scenario 
Fig.\ref{fig:phaseDG} (b), a large hysteresis occurring as a consequence of a 
large jump of 
magnetization, i.e., a hysteresis at thermal equilibrium resulting directly 
from the MF-FOT, cannot occur. According to the experimental data 
\cite{Movshovich,Tayama,Bianchi} in the situation of our interest in this 
paper corresponding to CeCoIn$_5$ under a field ${\bf H} \parallel c$, 
the phase diagram of Fig.\ref{fig:phaseDG} (a) seems to be always realized. 
However, it is 
theoretically valuable to clarify whether Fig.\ref{fig:phaseDG} (b) 
may be realized or not in 
real systems. Results of our numerical simulation, performed for 
examining our theoretical arguments given above, will be reported below.

\subsection{B. Simulation results}

\begin{figure}[b]
\scalebox{0.35}[0.35]{\includegraphics{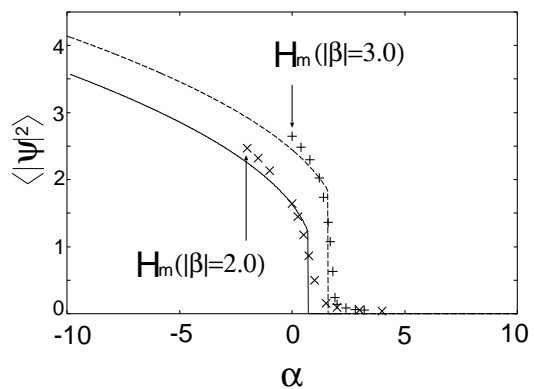}}
\caption{ Numerical data of $\alpha$-dependence (i.e., $H$-dependence) 
of $\langle |\Psi|^2 \rangle$ in 2D case, for $|\beta|=2.0$ (crossed symbols) and 
$|\beta|=3.0$ (plus symbols). 
Dashed and solid curves are the corresponding MF results. 
According to eq.(\ref{a0c}), the MF transition point corresponds to 
$\alpha=0.75$ $(1.69)$ in $\beta=2.0$ $(3.0)$. 
The system size ($6$,$4$) was commonly used. 
} 
\label{fig:psi2_2D}
\end{figure}

\begin{figure}[t]
\scalebox{0.8}[0.8]{\includegraphics{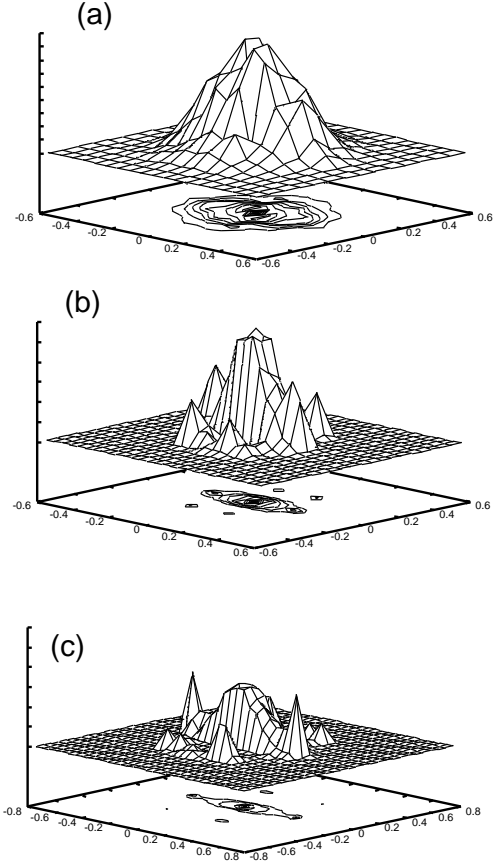}}
\caption{ $S({\bf k})$ data 
corresponding to the $|\beta|=2.0$ data in Fig.\ref{fig:psi2_2D} 
at (a) $\alpha=0.75$ and (b) $-2.0$.
(c) $S({\bf k})$ at $\alpha=-2.0$ for larger system size ($8$, $6$).} 
\label{fig:Sq_2D}
\end{figure}

In this subsection, we explain our methods and results of Monte Carlo (MC) simulations 
for the model eq.(\ref{statsum}). 
 Our simulation method closely follows that used in the simulations \cite{Kato,Hu} 
for the familiar GL model with a positive quartic term in place of $-|\beta|$ 
in eq.(\ref{Fsim}). On a fixed SC layer, the pair-field $\Psi$ is expanded in terms 
of the LLL basis functions $\phi_n(x,y)$ consistent with a quasi periodic 
boundary condition \cite{Kato} as $\Psi=\sum_n c_n \phi_n(x,y)$, 
and the system sizes $L_x$ and $L_y$ of a rectangular cell satisfy 
the commensurability with a triangular lattice ground state through the 
relation $L_x/L_y = \sqrt{3} N_x/2 N_y$ ($=2N_x/\sqrt{3}N_y$) in 2D (layered) case. 
Note that, as mentioned earlier, the ground state in the present case is, 
due to the neglect of nonlocality in the GL higher order terms, a triangular lattice, 
although the pairing state assumed originally is a four-fold $d$-wave one 
(see the sentences above eq.(\ref{Floc})). 
In the layered case, a periodic boundary condition is assumed across the 
layers. The system sizes we have studied were $(N_x$, $N_y)=(6$, $4)$ 
and $(8$, $6)$ in 2D case and $(6$, $6)$ in the layered case. 
The Markov chains for $c_n$ are generated by the Metropolis algorithm. 
For 2D case (layered case), we used $5 \times 10^5$ ($1.5 \times 10^6$) 
MC steps for thermalization and another 
$5 \times 10^5$ MC steps for further observation in both cases. 
As in the figures in Sec.II, 
we have assumed the value $0.8$ for 
$\mu_0 H_{c2}^{\rm orb}(0)/2 \pi T_{c0}$. 
Further, when assuming a $N(0)$ value appropriate to CeCoIn$_5$ \cite{SI} 
and that $T/T_{c0}=0.1$, the $|\beta(T)|$ value is estimated to be in the 
range between $2.0$ and $3.0$ used in the simulations.

To study fluctuation effects on the MF-FOT, the mean-squared average of 
the pair-field $\langle |\Psi|^2 \rangle$ was calculated. 
It corresponds to the magnetization 
when $a_0$ is the measure, primarily, of $H$. 
Hence, if it shows not a true discontinuous jump but a {\it rounded} behavior at 
$H_{c2}$ broadening with decreasing $|\beta|$, a genuine FOT at $H_{c2}$ is 
judged to be absent. Further, as a measure of the vortex-positional ordering 
(vortex-solidification), we have examined the structure factor $S({\bf k})$ 
defined \cite{RI90f} as the Fourier transform of the correlation function of 
$|\Psi({\bf r})|^2$. In the figures, $\alpha$-dependences of these physical 
quantities are shown. Although a change of $\alpha$ implies a change of $H$ 
at a fixed $T$, the vortex density is fixed at any simulation. 

\begin{figure}[b]
\scalebox{0.35}[0.35]{\includegraphics{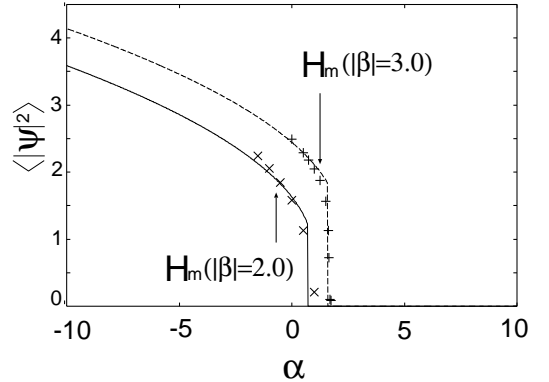}}
\caption{Numerical data of $\langle|\Psi|^2 \rangle$ in the layered 
case composed of four layers. 
The crossed symbols (plus symbols) are results in $|\beta|=2.0$ ($3.0$).} 
\label{fig:psi2_3D} 
\end{figure}

\begin{figure}[t]
\scalebox{0.4}[0.4]{\includegraphics{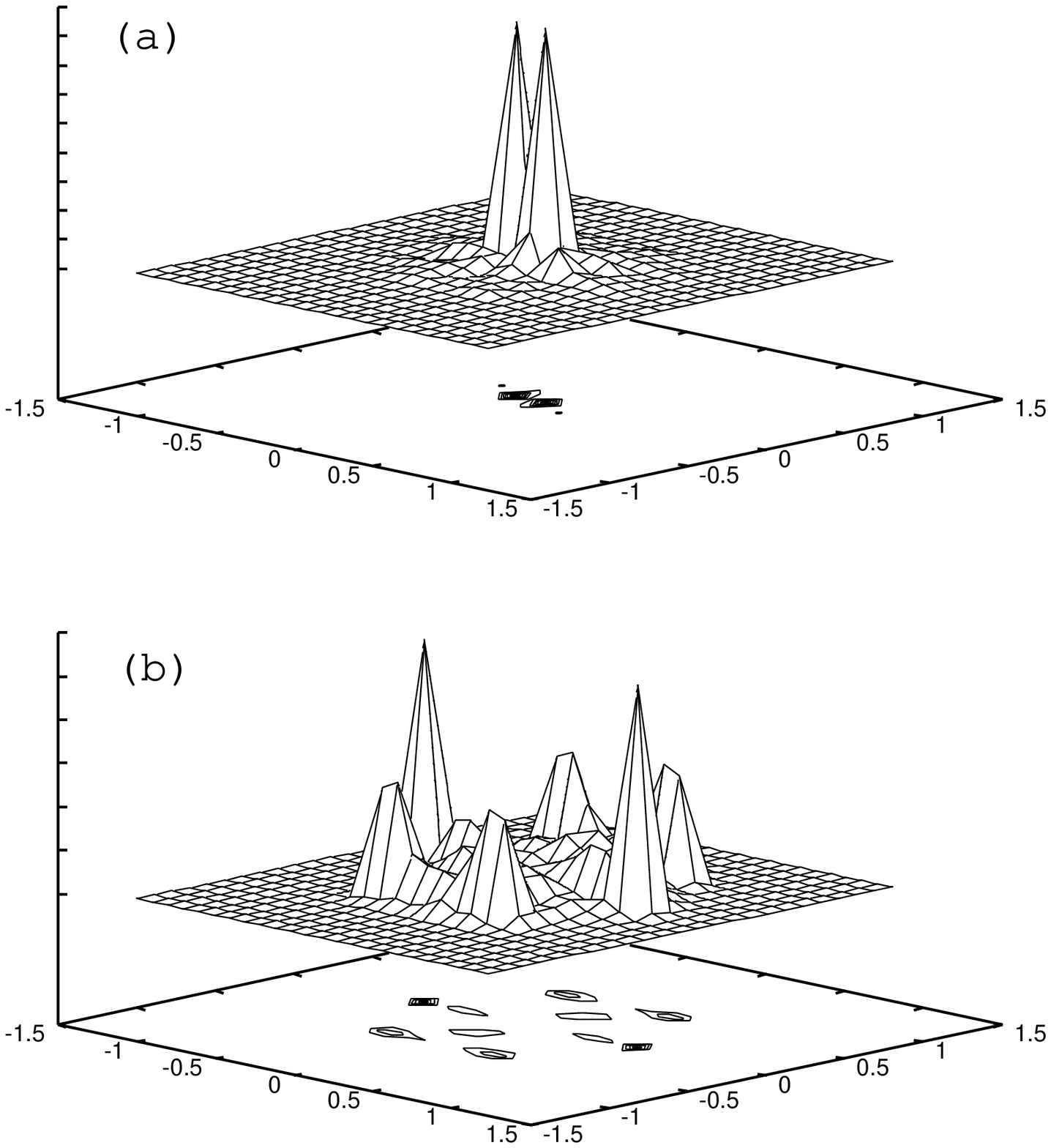}}
\caption{Structure factor defined from the correlation function of 
$|\Psi({\bf r})|^2$ for $|\beta|=3.0$ at (a)$\alpha=1.65$,(b) $\alpha=1.25$.}
\label{fig:Sq_3D} 
\end{figure}

First, let us present and explain 2D simulation results in which $\gamma=0$. 
The obtained results (symbols) are shown in Fig.\ref{fig:psi2_2D} and \ref{fig:Sq_2D}. 
The corresponding MF curves are also drawn for comparison.
The feature, that the simulation data lie above the MF curve in $H_m<H<H_{c2}$, 
is not surprising but presumably a reflection of a dimensionality dependence 
of the amplitude fluctuation (compare with Fig.\ref{fig:psi2_3D} below). 
It is found in the literature \cite{Ikeda91} that a similar dimensionality dependence 
appears in the magnetization in the case with a positive quartic coefficient. 
As is clear particularly from the $|\beta|=2$ data of Fig.\ref{fig:psi2_2D}, 
the nearly discontinuous jump of $\langle |\Psi|^2 \rangle$ 
at $H_{c2}$ is rounded due to the fluctuation, and thus, no genuine FOT has 
occurred at $H_{c2}$. 
We note that $\alpha_0(T = 0.1 T_{c0}) \simeq 50$ for 
the parameter values used here. Hence, if the abscissa in Fig.\ref{fig:psi2_2D} 
is reexpressed as the reduced field $(H-H_0)/H_0$, even this rounded behavior 
of $\langle |\Psi|^2 \rangle$ for $|\beta|=2$ 
cannot be distinguished from a strictly sharp 
discontinuity, and the difference $H_{c2}-H_m$ will not be visible (see Sec.IV). 
Thus, the presence of a genuine FOT cannot be argued through merely a steep 
growth of $\langle|\Psi|^2\rangle$ in a real system with weak SC fluctuation.

Next, let us examine whether the melting position coincides or not with 
the MF transition field $H_{c2}(T)$. The melting transition is widely 
believed to be a {\it weak} FOT, 
and this should be found~\cite{Kato} in Monte Carlo simulations 
as a tiny discontinuity in thermodynamic quantities. 
Unfortunately, due primarily to numerical difficulties, 
our simulation is restricted to too small systems to observe 
such a discontinuity. For our purpose of determining $H_m$, however, it is 
sufficient to find where the Bragg peaks of the vortex lattice disappear. 
Figure \ref{fig:Sq_2D} shows snapshots of the structure factor on 
$|\Psi({\bf r}_\perp)|^2$ when $|\beta|=2.0$. 
No vortex positional ordering is seen 
at $H_{c2}$. By comparing Fig.\ref{fig:Sq_2D} (a) and (b) with Fig.\ref{fig:psi2_2D}, 
one finds that 
nearly sharp Bragg peaks appear at $H_m$ 
below $H_{c2}$, while most of the entropy has been 
lost near $H_{c2}$ above it. The two field (or temperature) scales, 
one characterizing the steep growth of $\langle |\Psi|^2 \rangle$ 
and another corresponding to the sudden growth of vortex positional 
ordering, are clearly distinguished. 

We have also examined the size dependence of $S({\bf k})$ data. 
By comparing the data in Fig.\ref{fig:Sq_2D} (b) and (c) with each other, 
it will be clear that the six-fold symmetry of Bragg peaks is more remarkable 
in the former, i.e., in a smaller size. It means that 
the solidification is enhanced by the boundary condition in a smaller 
system. This size dependence will be sufficient for justifying our expectation that the vortex solidification (or melting) occurs below $H_{c2}$. 

\begin{figure}[t]
\scalebox{0.6}[0.6]{\includegraphics{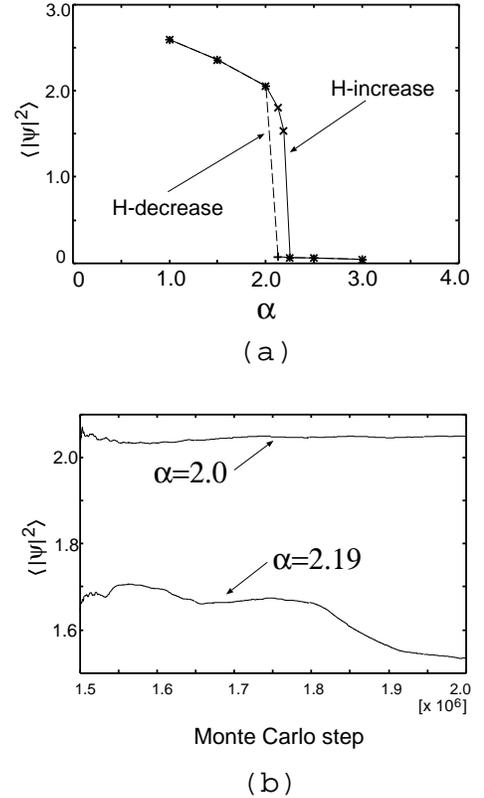}}
\caption{(a) Numerical data, similar to Fig.\ref{fig:psi2_3D}, for $|\beta|=3.5$. 
(b) The history of $\langle |\Psi|^2 \rangle$ 
at $\alpha=2.0$ and $\alpha=2.19$ in the $H$-increase process.
Note the strong dependence on Monte Carlo steps of 
$\langle |\Psi|^2 \rangle$ when $\alpha=2.19$, 
i.e., in the vicinity of $H_{c2}$. } \label{fig:3D_hys}
\end{figure}

\begin{figure}[t]
\scalebox{0.6}[0.6]{\includegraphics{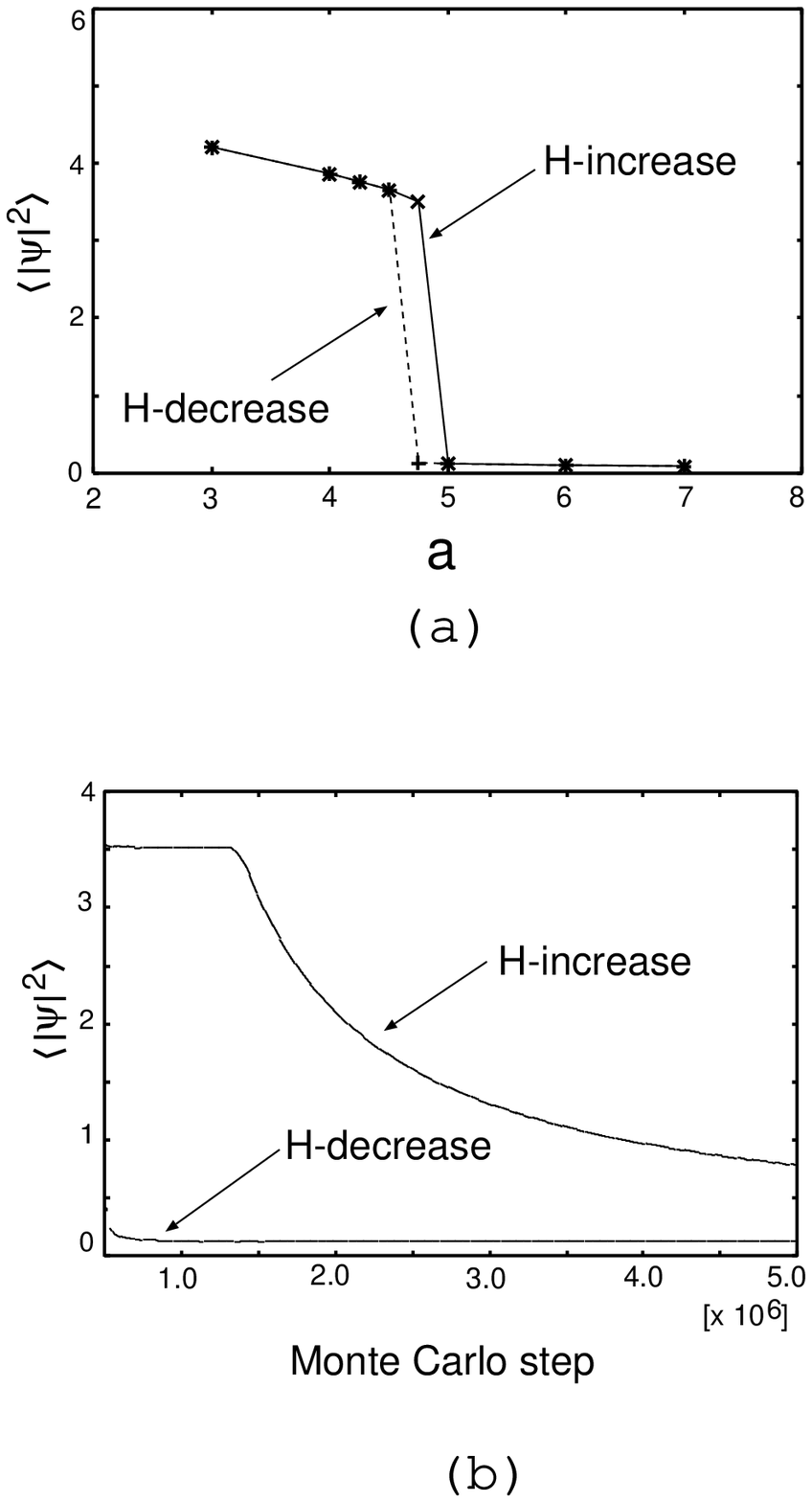}}
\caption{(a) $\langle |\Psi|^2 \rangle$ Data similar to Fig.\ref{fig:3D_hys} 
(a) for the 1D GL model, eq.(\ref{F1d}), taken at MC steps $1 \times 10^6$. 
(b) Data at $a=4.74$ showing a recovery of thermal equilibrium after many MC steps.}
\label{fig:1DGL}
\end{figure}

 The corresponding results in a layered (quasi 2D) 
system consisting of four layers are 
shown in Figs.\ref{fig:psi2_3D} and \ref{fig:Sq_3D}, 
where the parameter values $\gamma=0.25$ and $|\beta|=2.0$ or $3.0$ were assumed. 
The obtained computation results are essentially the same as in 2D case, 
except the feature that the difference $H_{c2}-H_m$ became narrower 
in the layered case. Thus, the anisotropy or dimensionality, i.e., 
the magnitude of $\gamma_0$, does not seem to induce an essential change of 
the true phase diagram. The nearly discontinuous behavior at $H_{c2}$ is also 
smeared out in this quasi 2D case as the fluctuation is enhanced, 
and no hysteresis is present accompanying this behavior at $H_{c2}$. 
The simulation results given above imply that, at least 
for $|\beta| \leq 3.0$, 
the correct phase diagram is Fig.\ref{fig:phaseDG} (a). 

Next, we report on consequences of an extension of simulation for the 
layered system composed of four layers to {\it weaker} fluctuation cases 
with $|\beta| > 3.0$ (i.e., at lower temperatures). 
As the numerical data in Fig.\ref{fig:3D_hys} (a) show, a hysteresis 
in the vicinity of $H_{c2}$ suggestive of a genuine FOT appears between 
two $\langle |\Psi|^2 \rangle$ curves for $|\beta|=3.5$, respectively, 
in increasing $H$ (corresponding to a heating) and decreasing $H$ 
(corresponding to a cooling). However, this hysteresis is {\it not} due to 
the vortex lattice freezing or melting included in the scenario of 
Fig.\ref{fig:phaseDG} (b) because, as the simulation results 
in $|\beta| \leq 3$ 
have shown, the hysteresis accompanying the melting is unobservably small 
(Although the melting field $H_m$ is estimated to lie close to 
$\alpha=2.0$ through $S({\bf k})$ data, it is not easy to 
conclude a separation between $H_m$ and $H_{c2}$ for this case). 
Further, since, as mentioned below eq.(\ref{coef}), an increase 
of $|\beta(T)|$ 
can be regarded as a decrease only of $T$ in the partition function 
(\ref{statsum}) under the same ${\cal F}_{\rm loc}$ (\ref{Floc}), 
it is difficult to imagine a scenario \cite{comdia} that the MF result would become exact 
in a very low but {\it finite} $T$ region. Actually, the hysteresis in 
Fig.\ref{fig:3D_hys} (a) is not due to a genuine FOT in thermal equilibrium: 
The $\alpha=2.19$ data in Fig.\ref{fig:3D_hys} (b) show that the system, at 
least 
in the vicinity of $H_{c2}$, has not reached the thermal equilibrium even 
during the MC steps we could observe.

Actually, a similar situation occurs, if the fluctuation is weak enough, 
in other systems with a MF-FOT but {\it no} true phase transition. 
To illustrate this statement, we show in Fig.\ref{fig:1DGL} results, 
corresponding to Fig.\ref{fig:3D_hys}, for a familiar 1D GL model with
 a negative quartic coefficient 
\begin{eqnarray}
{\cal F}_{1d} &=& \int dx \bigg( a |\Psi(x)|^2 
+ f \biggl|\frac{d \Psi(x)}{dx} \biggr|^2 
- \frac{|w|}{2} |\Psi(x)|^4 \nonumber \\ 
&&+ \frac{1}{3} |\Psi(x)|^6 \bigg), 
\label{F1d}
\end{eqnarray}
where $\Psi$ is a function only of $x$, and $a$ in this case may be regarded 
as a temperature variable. 
This 1D model is used here for comparison because the fluctuation in 3D GL model 
within LLL is expected to be similar to that of the corresponding 1D GL model 
in zero field. 
We chosen the values $|w|=5.0$ and $f=4.0$, together with the system size 
composed of $10^2$-sites. 
Due to a smaller number of degrees 
of freedom in SC fluctuations in this 1D case, a tendency of a full relaxation 
to the thermal equilibrium (i.e., a disappearance of hysteresis), 
as Fig.\ref{fig:1DGL} (b) shows, manages to be verified within the practically 
observable MC steps. In contrast, in the case of quasi 2D systems expressed 
by eq.(\ref{Fsim}), 
it is quite difficult to verify such a full relaxation within 
practically possible MC steps because of a much larger number of degrees 
of freedom in a quasi 2D vortex state.

Summarizing this subsection, for any $|\beta|$-value with $H_m$ lying below 
$H_{c2}$, no true FOT occurs at $H_{c2}$, and the phase diagram 
Fig.\ref{fig:phaseDG} (a) is justified. For larger $|\beta|$ values 
when $H_m$ may not be distinguished from $H_{c2}$, it is practically difficult 
to verify whether Fig.\ref{fig:phaseDG} (b), in place of Fig.\ref{fig:phaseDG} (a), 
is realized as a true phase diagram, and a hysteresis can appear 
in numerical experiments at $H_{c2}$ even without a genuine FOT occurring 
there.

\section{IV. Discussion}

As explained in Introduction, the present work was originally motivated as 
an extension of the problem of vortex phase diagram to the more general cases 
with spin (paramagnetic) depairing. Since the absence of MF second order transition 
at $H_{c2}$ in lower fields is well-established, it is unreasonable to expect 
the MF-FOT at $H_{c2}$ resulting from the Pauli paramagnetic depairing in higher 
fields to truly occur as a genuine FOT. The recent finding of an FOT-like 
nearly discontinuous crossover at $H_{c2}$ in the heavy fermion superconductor 
CeCoIn$_5$ in ${\bf H} \parallel c$ \cite{Tayama,Bianchi} provides us with 
a good occasion of a detailed comparison between the present theory and real data. 
Further, recent data in ${\bf H} \perp c$ \cite{Movshovich} showing 
a small hysteresis in heat capacity and suggesting a second order transition 
between the FFLO-like vortex solid and the ordinary vortex solid will stimulate 
detailed theoretical studies of vortex phase diagram in the region with 
paramagnetic depairing. 

As shown in Sec.II, 
in quasi 2D systems under fields perpendicular to the layers, 
a FOT at $H_{c2}$ should 
naturally occur in $T < T^*$ at the mean-field level 
in clean enough superconductors 
with a moderately strong paramagnetic depairing, 
and the onset temperature $T_{\rm FFLO}$ from which an FFLO-like vortex solid 
begins to appear lies much {\it below} $T^*$. 
The results, that this modulated solid state is rarely seen compared with 
the {\it apparent} FOT at $H_{c2}$ and is easily lost by a small 
amount of impurities, 
are consistent with the observations \cite{Movshovich} in a heavy fermion 
superconductor CeCoIn$_5$ in ${\bf H} \parallel c$ but seem to be a contrast 
to the conclusions on phase diagram in previous works \cite{Houzet}. 
Further, the range of $T^*$ ($0.2 < T^*/T_{c0} < 0.36$) shown 
in Fig.6 following from the value 
$\mu_0 H_{c2}^{\rm orb}(0)/2 \pi T_{c0} = 0.8$ and 
small $(\tau T_{c0})^{-1}$-values 
is comparable with that ($T^*/T_{c0} \simeq 0.25$) in CeCoIn$_5$ under ${\bf H} \parallel c$ \cite{Movshovich,Tayama,Bianchi}. 
When the $N(0)$-value in CeCoIn$_5$ in zero field \cite{SI} is used further, 
we find the nearly discontinuous jump value of magnetization at $H_{c2}$ 
when $T=50$ (mK) to become $20$ (G) which is comparable with the estimated 
value $30$ (G) in ref.\cite{Tayama}. 
Further, the width $\Delta H(T)$ of the magnetization jump 
at $H_{c2}$ will be 
estimated using the numerical data of Fig.\ref{fig:psi2_3D} 
as follows. 
Using the relations $\alpha_0(T) \simeq 50 (0.1 T_{c0}/T)^{2/3}$ 
and $\beta(T) \propto T^{-1/3}$, suggested in Sec.III, 
 and assuming $|\beta(T=0.1 T_{c0})|=2.0$, the data in Fig.\ref{fig:psi2_3D} 
imply $\Delta H(T=0.1 T_{c0}) \simeq 0.1$(T) and 
$\Delta H(T=0.025 T_{c0}) \simeq 0.03$ (T), which seem to be rather narrower 
than those of available data \cite{Tayama,Murphy}. 
It is not surprising because, even in a genuine FOT, an entropy 
(or a magnetization) jump is broadened due to an inhomogeneity in real materials. 

As shown and mentioned in Sec.III, 
the absence of a true FOT at $H_{c2}$ should be an appropriate interpretation 
for observations \cite{Tayama,Bianchi} at 
least in ${\bf H} \parallel c$ where no measurable hysteresis was 
observed although the ordinary magnetic hysteresis related to the peak 
effect near $H_{c2}$ \cite{V3SI} might appear. On the other hand, 
observations of a tiny hysteresis were recently reported 
in specific heat data \cite{Movshovich,Radovan} and also in magnetization data \cite{Tayama} of CeCoIn$_5$ in ${\bf H} \perp c$. 
Examining {\it microscopic} aspects in ${\bf H} \perp c$ leading to a MF phase diagram 
is beyond the scope of the present paper and will not be considered here. 
However, the issue of the genuine phase diagram in Sec.III 
is more generic and may be applicable to the ${\bf H} \perp c$ case in CeCoIn$_5$. Because the fluctuation effect in ${\bf H} \perp c$ is weaker than that in ${\bf H} \parallel c$ at the same $T$, the observed hysteresis 
in refs.\cite{Movshovich,Tayama,Radovan} can be understood within 
the present theory arguing the absence of a genuine FOT at $H_{c2}$, 
if it has the same origin as a hysteresis in numerical simulations at 
{\it low enough} $T$, shown in Figs.\ref{fig:3D_hys} and \ref{fig:1DGL}, 
which arises from an {\it incomplete} 
relaxation at long but finite time scales 
in a system with a strong MF-FOT. 
In relation to this, we point out that the onset of hysteresis 
in ref.\cite{Movshovich} lies slightly {\it above} the temperature $T_{\rm FFLO}$ 
at which $H_{\rm FFLO}(T)$ branches from the MF-FOT (i.e., $H_{c2}(T)$) line. 
It implies that the onset of hysteresis has nothing to do with the appearance 
of the FFLO-like state and thus that there is no physical reason favoring 
a termination of some genuine SC (superconducting) FOT. Alternatively, the observed hysteresis 
might accompany a true {\it magnetic} FOT induced by the $\langle |\Delta|^2 \rangle$
-nucreation 
at $H_{c2}$ as a consequence of a coupling between the SC and a magnetic 
order parameters. In relation to this, we point out the observation of 
a large magnetic hysteresis \cite{Murphy} in ${\bf H} \perp c$ just below $H_{c2}$ 
to which the corresponding one was not seen in the vanishing of resistivity 
(i.e., a SC ordering). 

 Finally, let us point out that the present theory easily explains why, 
in fields parallel to the layers, the transition to an FFLO-like phase and 
the nearly discontinuous crossover at $H_{c2}$ implying the MF-FOT were 
measured not in organic materials with larger anisotropy but in a heavy fermion 
material with weaker anisotropy. At least at the MF level, the case with 
a field parallel to the layers in more anisotropic materials is subject to 
a stronger paramagnetic depairing and is the best candidate for observing 
the paramagnetic effects such as the FFLO-like state and the MF-FOT. 
The organic materials satisfy this requirement, and actually, the upwardly 
increasing $H_{c2}(T)$ curve determined resistiviely \cite{Naughton} 
under high fields parallel to the layers is an evidence \cite{com3} that the spin depairing is microscopically important 
without being disturbing by an impurity effect. However, both the MF-FOT and 
a transition to an FFLO-like state have not been seen in the organic materials. 
On the other hand, since such a upward $H_{c2}(T)$-curve is not visible 
in the heavy fermion material CeCoIn$_5$ with a much weaker anisotropy, 
one might wonder why these crossover and transition arising from the spin 
depairing have occurred in this material. This puzzling facts are easily 
resolved by taking account of fluctuation effects examined in this paper. 
Typically, in the organic and cuprate materials \cite{Naughton,Shibauchi}, 
the fluctuation effect is much stronger compared with those of CeCoIn$_5$. 
Actually, a shorter coherence length tends to result in a larger Maki parameter 
and simultaneously to enhance the fluctuation even in the parallel field 
case \cite{Isotani,Ada2}. Consequently, as shown in Sec.III, 
the MF-FOT behavior is rounded and is transmuted into a broad continuous crossover 
by the SC fluctuation, reflecting the absence of the true FOT at $H_{c2}$. 
Further, a remarkable field and temperature range of the vortex liquid region 
in which the resistance is finite may be created below $H_{c2}(T)$-curve even 
in the parallel field case \cite{Ada2} where the fluctuation effect is minimized. 
Since the FFLO-like state is limited to a narrow field range below $H_{c2}$, 
and the modulation parallel to the field does not lead to any ordering 
in the vortex liquid, the vortex liquid region should mask and erase 
the FFLO-like phase in a strongly fluctuating superconductor. 
For these reasons, cleaner superconducting materials with {\it weaker} 
fluctuation such as CeCoIn$_5$ are the best candidates for examining 
the MF high field phase diagram in the case with Pauli paramagnetic effects.

\begin{acknowledgments}
We thank T. Sakakibara, Y. Matsuda, K. Machida, K. Izawa, T. Tayama, C. Capan, H. Radovan, and 
R. Movshovich for informative and stimulative discussions. 
Part of the numerical computation in this work has been carried out at the 
Supercomputer Center, Institute for Solid State Physics, University of Tokyo, 
and at the Yukawa Institute Computer Facility in Kyoto University. 
\end{acknowledgments}

\appendix

\section{Appendix A: derivation of $\hat{K}_2$}

 In this appendix we present how to solve the eigenvalue problem 
of $\hat{K}_2$ or equivalently of $\hat{D}$. 
 Using the identity $1/\alpha=\int_0^{\infty} d\rho e^{-\alpha \rho}$ 
and after the energy integration, we get the differential operator of 
infinite order 

\begin{eqnarray}
\hat{D}_d(2\varepsilon) &=&
 \sum_{\varepsilon > 0} 
 \int_{0}^{\infty} d \rho 
 e^{-(2\varepsilon+1/\tau) \rho} 
{\cal J}_0 \Big(2J\sin(\frac{q_z s}{2}) \rho \Big)\nonumber \\
&&\times 
\cos(2I\rho) \Big[ \left\langle 
|w_{\bf p}|^2 e^{-i{\bf v}\cdot \Pi \rho } \right\rangle_{\bf \hat{p}} 
 + c.c. \Big],  
\end{eqnarray}
 where ${\cal J}_0$ is the zeroth order Bessel function.
Expanding the exponential and averaging on the Fermi surface, 
we have 
\begin{eqnarray}
\left\langle |w_{\bf p}|^2 e^{-i{\bf v}\cdot \Pi \rho } \right\rangle_{\bf \hat{p}}
&=& e^{-(\frac{\rho}{2\tau_H})^2 }
\sum_m \frac{\Big( -2 \big( \frac{\rho}{2\tau_H}\big)^2 \Big)^m}{(m!)^2} 
\hat{\pi}_+^m \hat{\pi}_-^m \nonumber \\
&& \qquad + \quad {\rm off\quad diagonal\quad terms},
\end{eqnarray}
where  $\hat{\pi}_{\pm}$ are given below the eq. (17),  
and a circular Fermi surface was assumed. 
As explained above eq.(9), we have only to focus on the diagonal terms. 
Noticing the eigenvalue of $\hat{\pi}_+^m \hat{\pi}_-^m$ in the N-th Landau 
level 
is $\frac{N!}{(N-m)!}$ and performing the m-summation, 
we finally obtain eq. (9).

\section{Appendix B: expressions of $I_4$ and $I_6$}
 In this appendix, we derive the expressions of $I_4$ and $I_6$
in terms of a convenient expression for 
$e^{{\rm i} \rho {\bf v}\cdot {\Pi}} u_{0,k}({\bf r}_\perp)$. 
 If we denote the position on a (2-dimensional) Fermi surface by a 
complex number $v_F \zeta = v_x + {\rm i} v_y$ and 
define $\lambda=\rho \zeta^*/\tau_H$, we have 
\begin{eqnarray}
e^{{\rm i} \rho {\bf v}\cdot{\bf \Pi}} 
&=&
e^{\frac{\rm i}{\sqrt{2}} (\lambda \hat{\pi}_+ + \lambda^* \hat{\pi}_-) }\nonumber \\
&=&
e^{-\frac{|\lambda|^2}{4}}e^{\frac{\rm i}{\sqrt{2}}\lambda \hat{\pi}_+}
e^{\frac{\rm i}{\sqrt{2}}\lambda^* \hat{\pi}_- }, 
\end{eqnarray}
where we used the operator identity 
$e^{\hat{A}+\hat{B}}=e^{-\frac{1}{2}[A,B]}e^{\hat{A}}e^{\hat{B}}$ valid when 
$[\hat{A},\hat{B}]$ is a constant.
 From this expression it is sufficient to know the form of 
$e^{\frac{\rm i}{\sqrt{2}}\lambda \hat{\pi}_+} u_{0,k}({\bf r}_\perp)$.  
 Using the above identity to obtain 
\begin{equation}
e^{\frac{\rm i}{\sqrt{2}}\lambda \hat{\pi}_+}
=
e^{-\frac{\lambda^2}{8}} e^{\frac{\lambda}{2} ({\rm i}r_H\partial_y-x r_H^{-1})}
e^{\frac{\lambda}{2}r_H \partial_x},  
\end{equation}
and noticing $e^{\alpha \partial_x} g(x)=g(x+\alpha)$ for any 
non-singular function $g(x)$, we finally have 
\begin{eqnarray}
e^{{\rm i} \rho {\bf v \cdot \Pi}} u_{0,k}({\bf r}_\perp)
&=&
e^{-\frac{1}{4}(|\lambda|^2-\lambda^2)}
e^{-\frac{1}{2}(x/r_H+kr_H+\lambda )^2+{\rm i}ky}. \nonumber \\
\\
e^{{\rm i} \rho {\bf v \cdot \Pi}^*} u_{0,k}^*({\bf r}_\perp)
&=&
e^{-\frac{1}{4}(|\lambda|^2-\lambda^{*2})}
e^{-\frac{1}{2}(x/r_H+kr_H - \lambda^* )^2-{\rm i}ky}. \nonumber \\
\label{eq:coherent}
\end{eqnarray}

\begin{widetext}
With the help of the above identities, the following results are 
easily derived. 
\begin{eqnarray}
&&
\left. 
\int \frac{d^2 r_{\perp}}{S_H}
e^{{\rm i} (\rho_1 {\bf v \cdot \Pi}_1^*
+\rho_2 {\bf v \cdot \Pi}_2
+\rho_3 {\bf v \cdot \Pi}_3^*
+\rho_4 {\bf v \cdot \Pi}_4 )} 
u_{0,k_1}^*({\bf r}_{\perp1})
u_{0,k_2}({\bf r}_{\perp2})
u_{0,k_3}^*({\bf r}_{\perp3})
u_{0,k_4}({\bf r}_{\perp4}) \right|_{{\bf r}_{\perp,i} \to {\bf r}_{\perp}}
\nonumber \\
&&\hspace{3cm}=
\frac{1}{\sqrt2} \delta_{k_1+k_3,k_2+k_4} e^{-\frac{r_H^2}{4}(k_{13}^2+k_{24}^2)}
e^{-\frac{1}{4}
\Big[ \sum_{i=1}^4 |\lambda_i|^2+\frac{1}{2}(\lambda_{13}^{*2}+\lambda_{24}^2)
+(\lambda_1^*+\lambda_3^*)(\lambda_2+\lambda_4)
+2r_H(k_{13}\lambda_{13}^*-k_{24}\lambda_{24}) \Big]} \nonumber \\
&&\hspace{3cm} \equiv \frac{1}{\sqrt{2}} \delta_{k_1+k_3,k_2+k_4}
e^{-\frac{r_H^2}{4}(k_{13}^2+k_{24}^2)} 
I_4(\{ \lambda_i\}) \label{eq:I4}
\end{eqnarray}

\begin{eqnarray}
&&
\left. 
\int \frac{d^2 r_{\perp}}{S_H}
e^{{\rm i} (\sum_{i:{\rm odd}}\rho_i {\bf v \cdot \Pi}_i^*
+\sum_{i:{\rm even}}\rho_i {\bf v \cdot \Pi}_i)
}
u_{0,k_1}^*({\bf r}_{\perp1})
u_{0,k_2}({\bf r}_{\perp2})
u_{0,k_3}^*({\bf r}_{\perp3})
u_{0,k_4}({\bf r}_{\perp4}) 
u_{0,k_5}^*({\bf r}_{\perp5})
u_{0,k_6}({\bf r}_{\perp6}) 
\right|_{{\bf r}_{\perp,i} \to {\bf r}_{\perp}}
\nonumber \\
&&=
\frac{1}{\sqrt{3}} \delta_{k_1+k_3+k_5,k_2+k_4+k_6}
e^{-\frac{r_H^2}{6}\sum_{(i,j)}k_{ij}^2} \nonumber  \\ 
&&\times e^{
-\frac{1}{4}\Sigma_{i=1}^6 |\lambda_i|^2 
+\frac{1}{4}({\Sigma \atop {i:{\rm odd}}}\lambda_i^{*2} 
+{\Sigma \atop {\tiny i:{\rm even}} } \lambda_{i}^2)
-\frac{1}{12}({\Sigma \atop {i:{\rm odd}} }\lambda_i^{*}
+{\Sigma \atop {i:{\rm even}} }\lambda_{i})^2
-\frac{1}{6}({\Sigma \atop{(i,j):{\rm odd}} }\lambda_{ij}^{*2}
+{\Sigma \atop {(i,j):{\rm even}}} \lambda_{ij}^2)
+\frac{r_H}{3}({\Sigma \atop {(i,j):{\rm odd}}} k_{ij}\lambda_{ij}^{*}
-{\Sigma \atop {(i,j):{\rm even}}} k_{ij}\lambda_{ij}
)} \nonumber \\
&\equiv& \frac{1}{\sqrt{3}} \delta_{k_1+k_3+k_5,k_2+k_4+k_6}
e^{-\frac{r_H^2}{6}\sum_{(i,j)}k_{ij}^2}
I_6(\{ \lambda_i\})\label{eq:I6}
\end{eqnarray}
where $(i,j)=\{(1,3),(3,5),(5,1),(2,4),(4,6),(6,2) \}$. 

\end{widetext}


\begin{thebibliography}{99}
\bibitem{comdia} One might imagine a possibility that the MF-FOT, 
as in the classical liquid-gas transition, occur above $H_m$ as a real FOT 
with no symmetry breaking accompanied. 
However, this possibility is certainly excluded by our simulation results 
given above and will not be considered hereafter as another possibility.
\end{thebibliography}

\end{document}